\documentclass[sigconf]{acmart}

\AtBeginDocument{%
  \providecommand\BibTeX{{%
    \normalfont B\kern-0.5em{\scshape i\kern-0.25em b}\kern-0.8em\TeX}}}

\usepackage{enumitem}
\usepackage{soul}
\usepackage{float}
\usepackage{xcolor}
\usepackage{subfig}
\usepackage{cancel}

\begin{document}

\title[UPI's Impact on Spending Behavior Among Indian Users and Prototyping Financially Responsible Interfaces]{From Cash to Cashless: UPI's Impact on Spending Behavior Among Indian Users and Prototyping Financially Responsible Interfaces}

\author{Harshal Dev}
\email{harshal19306@iiitd.ac.in}
\affiliation{%
  \institution{IIIT Delhi}
  \city{New Delhi}
  \country{India}
}

\author{Raj Gupta}
\email{raj21410@iiitd.ac.in}
\affiliation{%
  \institution{IIIT Delhi}
  \city{New Delhi}
  \country{India}
}

\author{Sahiti Dharmavaram}
\email{sahiti.dharmavaram2021@vitstudent.ac.in}
\affiliation{%
  \institution{VIT Vellore}
  \city{Vellore}
  \country{India}
}

\author{Dhruv Kumar}
\email{dhruv.kumar@iiitd.ac.in}
\affiliation{%
  \institution{IIIT Delhi and BITS Pilani}
  \city{New Delhi}
  \country{India}
}



\begin{abstract}
Unified Payments Interface (UPI) is a groundbreaking innovation making waves in digital payment systems in India. It has revolutionised financial transactions by offering enhanced convenience and security. While previous research has primarily focused on the macroeconomic effects of digital payments, our study examines UPI's impact on individual spending behavior. Through a survey of 276 respondents and 20 follow-up interviews, we found that approximately 75\% of participants reported increased spending due to UPI. Many attributed this to UPI’s intangible nature, which reduced feelings of guilt typically associated with spending. Additionally, participants provided suggestions to improve the user experience of existing UPI applications. Utilizing this feedback, we developed a high-fidelity prototype based on a popular UPI app in India and conducted usability testing with 34 participants. The insights gathered from this testing shaped the final prototype and its features. This study offers valuable design recommendations for UPI app developers and other stakeholders.

\end{abstract}



\received{20 February 2007}
\received[revised]{12 March 2009}
\received[accepted]{5 June 2009}

\maketitle

\section{Introduction}\label{sec:intro}
Digital payments have become integral to modern financial landscapes, offering swift, secure, and convenient electronic transactions that transcend traditional methods. This global shift towards digital transactions has been accelerated by the widespread adoption of internet and mobile technologies, reshaping financial practices across the world \cite{gulf_countries}. In 2022 alone, the volume of real-time digital transactions reached an unprecedented 195 million \cite{89.5_million}. 



India, in particular, has emerged as a global leader in digital transactions, accounting for 46 \% of the total global digital transactions in 2022 \cite{89.5_million}. Unified Payments Interface (UPI), launched in 2016 by the National Payments Corporation of India (NPCI),  has redefined financial transactions by enabling instant, real-time inter-bank transfers \cite{riseofupi}. As of August 2023, UPI has facilitated a staggering 10 billion transactions, totaling Rs 18,22,949.42 Crore(189 billion USD) \cite{npci_livemembers}. In India, 84 \% of all digital transactions in 2022, amounting to 1.79 trillion dollars, were channeled through UPI \cite{npci_productstatistics}. UPI's rapid proliferation and high penetration in the Indian payment ecosystem underscore its transformative impact \cite{riseofupi}.

Despite UPI's monumental success, a critical research gap exists regarding its influence on users' spending habits. Our study addresses this crucial gap in existing research on UPI, specifically focusing on analyzing the impact of UPI on users' spending habits. 
While previous studies \cite{cashless_economy_growth} acknowledge UPI's success in reshaping digital payments, they fall short of providing detailed insights into how this technology influences "individual" spending behaviors. The existing literature often emphasizes macro-level implications and adoption rates. Multiple questions about changes in expenditure preferences, budgeting strategies, 
and overall financial behaviors remain unexplored, hindering a holistic understanding of UPI adoption's intended and unintended consequences on users' daily financial practices. This study seeks to fill these critical gaps by providing in-depth insights into the micro-level effects of UPI on spending behaviors. 

Moreover, there’s a notable absence of research delving into potential variations in UPI’s impact on spending behaviour across different demographic groups like age, profession, socio-economic status, etc. Our study aims to bridge this gap as well. We also explore users' perceptions of security and how trust in UPI influences expenditure patterns, unraveling the intricate dynamics between trust, technology adoption, and financial decision-making. Additionally, we investigate the relationship between users' financial management levels and UPI utilization, shedding light on how understanding financial principles shapes spending choices. 

\noindent\textbf{Phase 1:} The above mentioned portion of the study constitutes \textbf{Phase 1}. In this phase, we employed a comprehensive mixed-methods approach to address the identified gaps, integrating qualitative and quantitative analyses. We began by distributing an initial survey to a diverse demographic to assess UPI usage, which resulted in 276 total responses, with 235 being valid. To gain deeper insights into user behavior, pain points, and preferences, we conducted semi-structured interviews with 20 respondents from the survey.

\begin{figure*}[htb]
    \centering
    \includegraphics[width=0.5\linewidth]{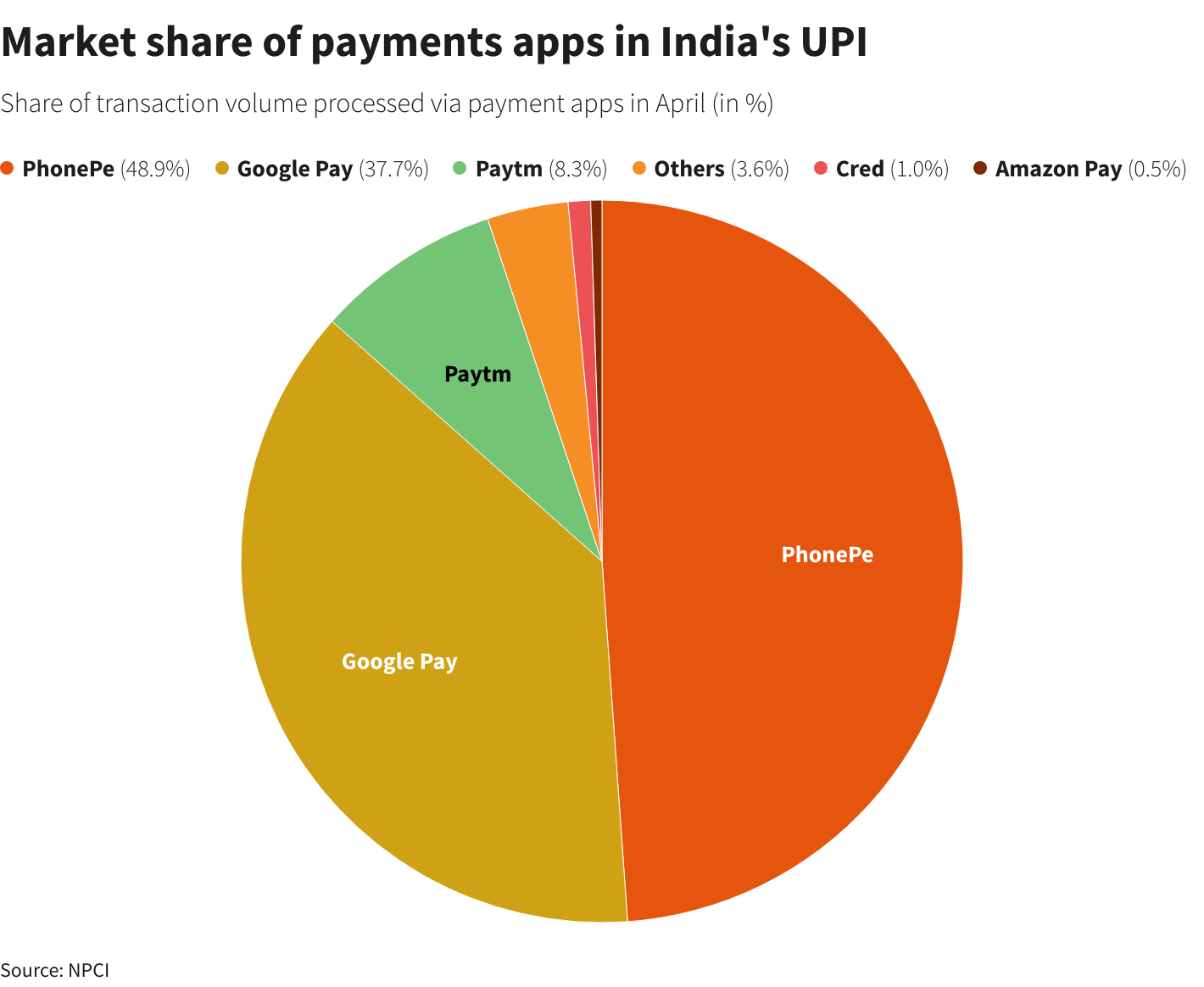}
    \caption{MARKET SHARE OF UPI PAYMENT APPS IN INDIA}
    \label{fig:marketshare}
\end{figure*}

\noindent\textbf{Phase 2:} Based on the feedback and recommendations gathered in \textbf{Phase 1}, 
we developed an initial prototype based on a popular UPI app in India (PhonePe \cite{phone_pe,phonepetimesofindia}) in \textbf{Phase 2}. We selected PhonePe as the base UPI app due to its significant 48.9\% market share, making it the largest UPI platform used in India (See Figure \ref{fig:marketshare}). To finalize the requirements for the initial prototype, we also analyzed the existing budgeting applications, employing a competitive lens to understand their impact on users' awareness of spending patterns and financial discipline. The developed prototype incorporated several new features, including an expense tracker, balance display, and budgeting limits, aimed at improving financial awareness and promoting responsible spending as detailed in \S\ref{sec:features}.

We then conducted usability testing on the prototype, involving 1-on-1, multi-stage personalized sessions with 34 participants. These sessions aimed to validate the prototype and gather detailed user feedback on new features and overall user experience. The usability testing provided valuable insights into users' preferences, areas of challenge, and further improvements. This iterative process led to the development of a refined prototype that not only met users' needs but also incorporated the most-requested features, with a focus on promoting responsible financial behavior and enhancing overall usability.

 This study aims to address the following research questions (\textbf{RQs}):
\begin{enumerate}
    \item \textbf{RQ1} \label{RQ1}: How does the introduction of UPI influence Indian users' spending habits, and what are the observed changes: increased, decreased, or unchanged spending patterns?
    \item \textbf{RQ2} \label{RQ2}: How do users perceive the convenience and efficiency of UPI transactions compared to traditional cash transactions, and to what extent do these perceptions shape their spending behavior? 
    \item \textbf{RQ3} \label{RQ3}: To what extent does UPI adoption vary across demographic groups, such as age and profession, and how do these variations influence users' financial practices, including spending habits and long-term financial goals?
    \item \textbf{RQ4} \label{RQ4}: What are the strengths and limitations of current UPI apps from a user perspective, and how can their design be refined to improve overall user satisfaction?
    
\end{enumerate}


Among the participants in \textbf{Phase 1}, \textbf{74.2\% reported increased spending due to UPI}, while only 7\% noted a decrease in spending. Notably, \textbf{91.5\%} of respondents \textbf{expressed satisfaction} with their UPI usage, and \textbf{95.2\% found making payments via UPI to be convenient}.

In \textbf{Phase 2}, based on the usability testing of the initial prototype, \textbf{87.9\% of users indicated that adding an expense tracker to financial apps like PhonePe \cite{phonepe} and Paytm \cite{paytm} would be beneficial.} This would provide them the ability to view expenses by month and category (e.g., food, travel), with a day-wise breakdown and budget allocation for each component. Additionally, \textbf{78.7\% of users agreed that seeing their balance before and after making a payment would be advantageous.}

This study offers valuable insights for HCI researchers, financial institutions, and policymakers, providing guidance for the development of digital payment systems that align with users' financial needs, behaviors, and promote responsible financial management. To the best of our knowledge, this is the first user study in India that explores UPI’s impact on individual spending behavior and prototypes UPI interfaces designed for financial responsibility.

\section{Related Work}\label{sec:related_work}


Mobile payment systems are progressively influencing people's socio-economic lives in South Asia \cite{chineseqr_2,chineseqr_3,chineseqr_24,chineseqr,chineseqr_52,chineseqr_97}. India topped the list of the five countries, namely Brazil, China, Thailand, and South Korea, for digital payments by the end of 2022 with 89.5 million transactions \cite{89.5_million}. The sheer nature of mobile payments and their ubiquity, along with convenience, helps include individuals, eventually benefiting the country's economy \cite{kenya,chineseqr_24}.



Historically, India has been characterized by high cash usage, with a substantial portion of the workforce engaged in the informal sector relying on cash transactions \cite{Misleadingdichotomy, digitalDiscontent}. The transformation from a cash-driven economy to a significant shift towards digital payments is also driven by various significant events like government's "Digital India" vision, Demonetization \cite{terrorism,chineseqr_68, Cash-Digital-Payments-and-Accessibility-A-Case-Study-from-India} and COVID-19 pandemic. According to the World Bank, over 80 million Indian adults made their first digital merchant payment after the pandemic's onset\cite{worldbank}. Studies have indicated the positive impact of electronic transactions on trade and economic development in India \cite{Abrazhevich, mukhopadyay}. The benefits include transparency, accountability, and reduced cash-related fraud, contributing to overall economic growth \cite{sudheer}. The surge of digital wallets and UPI platforms, including BHIM UPI \cite{bhim}, Paytm \cite{paytm}, PhonePe \cite{phone_pe}, and Google Pay \cite{google_pay}, has been instrumental in reshaping India's financial landscape. Internet penetration in the country rose from 14\% in 2014 to over 46\% in 2021, driven by affordable data tariffs introduced by regional mobile operators in 2016. This contributed to India becoming the second-largest global user base of the internet by 2018 \cite{2NDLARGEST,Jio}.

Despite progress, challenges remain, as only 35\% of Indian adults use bank accounts for digital payments \cite{globalfindex}. The low percentage (approx 5\% of the Indian population) of income tax contributors emphasizes the potential role of digital payments in increasing tax collection, as highlighted by previous research \cite{baxter}. The extensive penetration of smartphones in India, with 1.2 billion active mobile phone users, underscores the technological underpinnings of this digital transition \cite{MINT}. In 2016, a study by Goel \cite{goel} found that technological advancements have played a significant role in empowering and introducing innovations in E-payment systems.



Research in the field has extensively explored the impact of mobile payment systems on economies and societies, focusing on their transformative effects. Notably, people from different countries and different strata of society manage finances differently \cite{couplesharemoney,moneymatters,oldies}. This motivated us to understand the possible changes in their spending due to UPI and how Indian users keep track of their finances. 
This study addresses the micro-level effects of UPI on spending behaviors, filling gaps in existing literature focused on macro-level implications. It explores changes in expenditure preferences, budgeting strategies, and long-term financial goals. The research aims to comprehensively understand the consequences of UPI adoption on users' spending behavior while highlighting user concerns and preferences that influence the widespread acceptance of mobile payment systems in the region. The study also aims to provide insights beyond transaction numbers and uncover the factors that encourage users to trust the UPI technology, considering the backdrop of rising scams \cite{scam, scam_newspaper}.

Evidence suggests that individuals often under-save, fail to invest wisely, and are frequently indebted \cite{student_debt}. Several studies indicate that a significant fraction of individuals has a low level of financial knowledge, associated with sub-optimal financial outcomes in various areas such as retirement planning \cite{mitchell_lusardi,Lusardo_Tufano,rooij,llm}. Increasing financial literacy improves financial decision-making, leading governments worldwide to recognize it as a critical life skill, launching financial education initiatives to help young people acquire this skill \cite{asic, rbi-nsfe}.
Multiple studies have highlighted the importance of financial inclusion and literacy, with one study pointing out how roadblocks to digital or financial literacy impede digital financial inclusion \cite{financial_inclusion}. 
While cashless transactions promote technology literacy \cite{tanmay}, there is a visible gap in understanding how or if it promotes financial literacy. 
In contrast to the existing studies, our study aims to explore how financial literacy impacts spending habits, especially with the onset of cashless transactions (UPI) in the context of India.  



The use of UPI in India has been on the rise, but at the same time the number of UPI related scams have also been increasing. The massive growth of UPI, however, has has attracted the attention of fraudsters, who are increasingly targeting vulnerable users. According to Indian government figures, there were more than 95,000 cases of fraud involving UPI in the financial year ending April 2023, up from 77,000 in the previous year \cite{scammers}. According to Dixit et al. \cite{businesstoday} , the reason why there is a surge in number of UPI scams is not just the growing popularity of digital payments, but also a lack of financial literacy and imprudent use of technology, rendering a vast population vulnerable to such attacks. 
For instance, Allums et al. \cite{experience} underscores the importance of creating frictionless interactions and ensuring robust data protection to meet user expectations, particularly in payment scenarios . Surendran et al. \cite{ieee} focuses on the need for simplifying e-payment apps to improve adoption among users with limited technology access and low literacy, using apps like WhatsApp as examples of effective simplification . Gupta et al. \cite{elderly} emphasizes the development of secure e-wallet authentication designs, incorporating device-specific identity features and a two-phase authentication process to enhance security and reliability . Additionally, Hassan et al. \cite{electronics}on mobile expense tracking applications aims to create user-friendly solutions for young adults, helping them manage and visualize their daily transactions and budgeting . Finally, Kiat et al. \cite{cashsave} on prototype design and recommendations for e-wallet apps discusses enhancing inclusivity but does not delve deeply into the iterative design process or how user feedback integrates into refining prototypes. 

 Our research fills this gap enhances app functionality by grounding the prototype development in iterative testing and real-world feedback. It significantly boosts user satisfaction, ensuring the final product is practical and user-centric.

\section{Overview of the Study}\label{sec:03_Overview_of_the_study .tex}
\begin{figure*}[htb]
    \centering
    \includegraphics[width=1\linewidth]{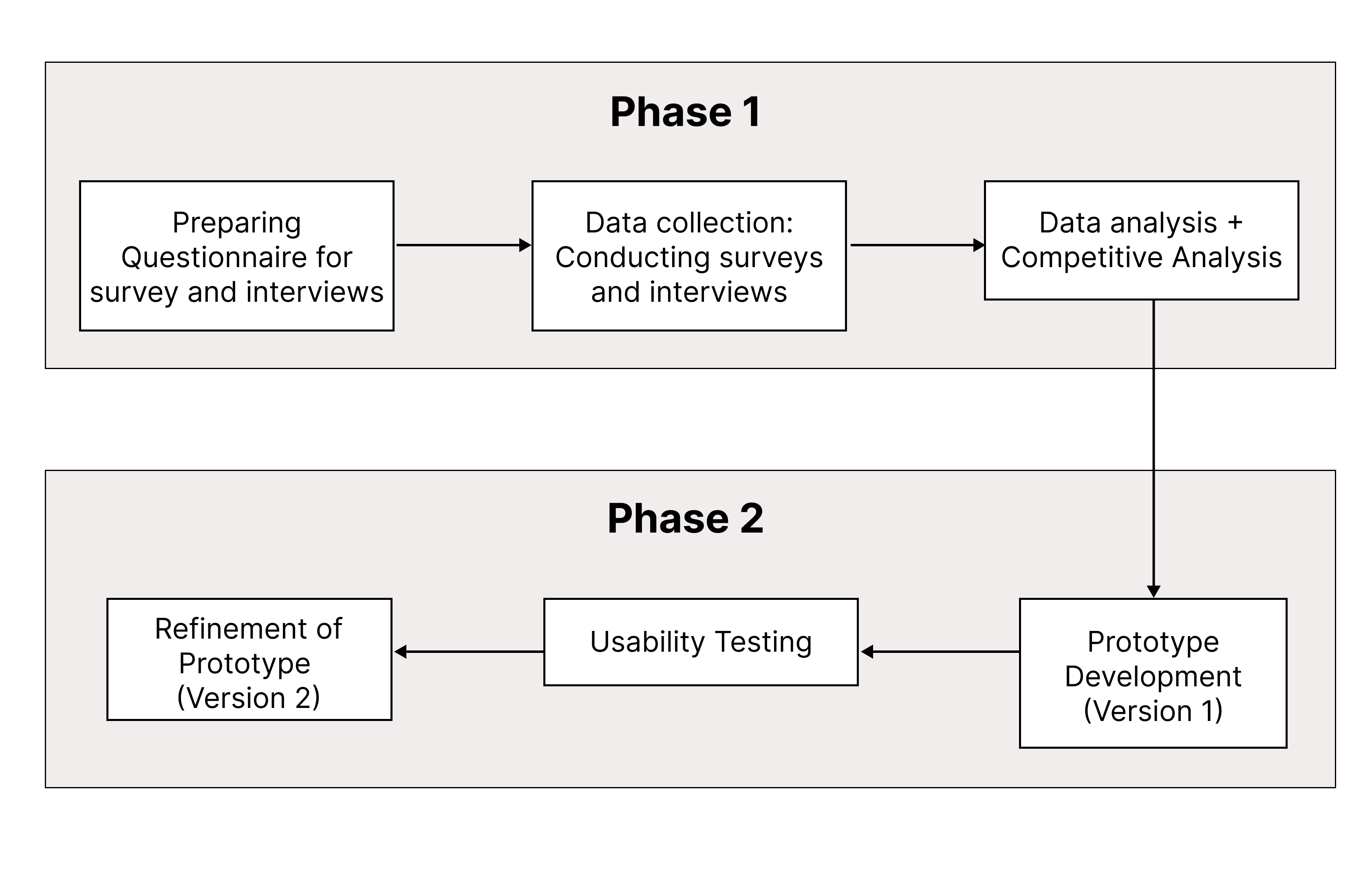}
    \caption{Overview of the entire research study}
    \label{fig:overview}
\end{figure*}

This section provides an overview of our study on the impact of UPI on spending behavior and the development of financially responsible UPI interfaces (see Figure \ref{fig:overview}). 

\noindent\textbf{Phase 1:} In \S\ref{sec:methods}, we outline the methodology used to collect data on UPI's influence on users' spending patterns. \S\ref{sec:evaluation} presents the results of both quantitative and qualitative analyses of the collected data. Finally, \S\ref{sec:Requirements_ for_next_gen.tex} summarizes the key insights from our data analysis and provides a competitive analysis of existing personal finance tracking apps.

\noindent\textbf{Phase 2:} Building on the insights from \textbf{Phase 1} regarding UPI’s impact on spending behavior and identifying key gaps, \textbf{Phase 2} focuses on prototype development and its usability testing. First, we developed the \textbf{initial prototype (version 1)} whose details are presented in \S\ref{sec:phase-2-initial-prototype}. \S\ref{sec:phase-2-methodology} presents the details of the usability testing of the developed prototype. The results of this usability study are presented in \S\ref{sec:results-usability-testing}. In \S\ref{sec:incorportating-user-recommendations}, we address the user feedback and recommendations gathered during usability testing, which informed the development of the \textbf{refined final prototype (version 2)}. Based on these findings, \S\ref{sec:recommendations-for-developers} provides recommendations for UPI app developers and stakeholders, focusing on improving user engagement and enhancing financial awareness. Lastly, \S\ref{sec:discuss} revisits the research questions introduced in the beginning and offers answers based on the study's findings. We conclude in \S\ref{sec:conclusion}.

\section{Phase 1 : Methodology}\label{sec:methods}

\subsection{Data Collection}
We employed a mixed-methods approach to investigate the impact of UPI on individual spending behaviors and financial practices. We utilized two primary data collection techniques: surveys and interviews. We used a random sampling method for the survey component, targeting individuals aged 18 and above. Our survey was conducted through Google Forms to ensure easy accessibility and respondent anonymity (by not asking for their personal information like name, phone number, etc). It garnered 276 responses from a diverse participant pool spanning various age groups and occupations. To ensure the quality, consistency and relevance of the responses, we implemented stringent criteria for validity. Frivolous responses containing irrelevant information were excluded, for instance, in the question regarding the initiation of UPI usage, any response indicating usage before 2016 was deemed invalid, considering that UPI was publicly launched that year. This validation process resulted in 235 valid responses. We conducted in-depth semi-structured interviews by randomly selecting interviewees from our survey respondents who expressed an interest in participating. We conducted 20 (13 male \& 7 female) interviews after receiving explicit consent from each of the participants. Survey and interview questions are presented in Appendix \ref{appendix:phase_1_survey} and Appendix \ref{appendix:phase_1_interviews}, respectively.

\subsection{Data Analysis}

The data collected in \textbf{Phase 1} was analyzed using both quantitative and qualitative methods, reflecting the mixed-methods approach employed in this phase. For the \textbf{quantitative analysis}, we aggregated and analyzed the survey data to identify overall trends and patterns in UPI usage and its impact on spending behavior. Key metrics such as the percentage of users reporting increased or decreased spending, satisfaction levels with UPI, and perceptions of convenience were derived from the survey responses. These aggregate numbers highlight the broader behavioral trends among UPI users across different demographics.

For the \textbf{qualitative analysis}, we conducted a thematic analysis of the interview transcripts. Using an inductive approach, we identified recurring themes and patterns related to user experiences, perceptions, and financial practices. This method allowed us to extract meaningful insights directly from the data without imposing pre-determined categories. Themes such as spending habits, security concerns, budgeting strategies, and recommendations for improving UPI applications were coded and categorized. The detailed thematic analysis provided a deeper understanding of individual perspectives and behaviors, complementing the broader trends observed in the survey data.

\subsection{Ethical Considerations} 
\label{sec:p1ethical}

In conducting this research, we implemented various ethical considerations to uphold transparency and safeguard the privacy and well-being of participants. Our study’s research materials and protocols underwent thorough review and approval by our university’s Institutional Review Board (IRB). Before participating in this study, all candidates received information about the study’s purpose, the voluntary nature of their involvement, and the assurance of anonymity
and confidentiality. 
Explicit consent was obtained from participants to record the interviews.

\subsection{
Limitations}
\label{sec:p1limit}

The research exclusively focused on
respondents from urban areas in India, where technology adoption
and usage, including electronic payment methods like UPI, tend
to be higher and well integrated into daily life. As a result, the
findings may not encompass the diverse spending habits. They may
not accurately represent the impact of UPI on the spending habits
of those living in rural areas in India.

\section{Phase 1: Evaluating the Impact of UPI on Spending Behaviour}\label{sec:evaluation}
\subsection{Quantitative Analysis (Analysis of Survey Results)}
\begin{figure}[h!]
  \centering
  \subfloat[]{\includegraphics[width=0.4\textwidth]{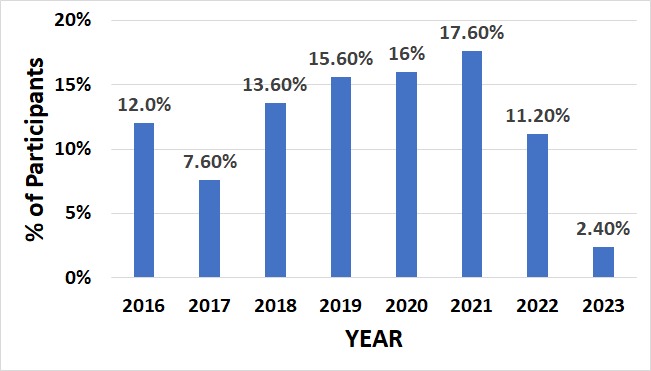}}\hspace{0.05\textwidth}
  \subfloat[]{\includegraphics[width=0.4\textwidth]{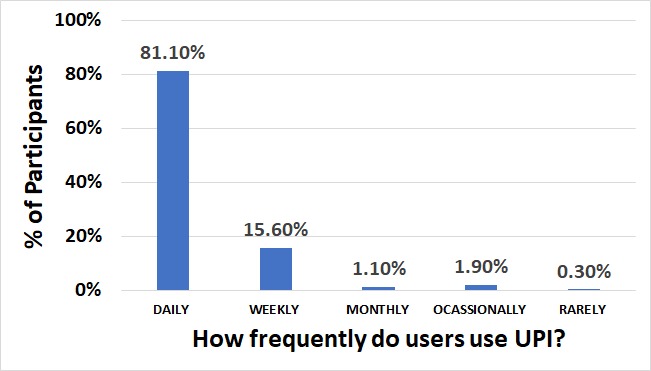}}\hspace{0.05\textwidth}
  \caption{(a) When did users start using UPI?  (b) How often do users use UPI? }
  \label{fig:figures_row1}
  
\end{figure}

Our survey received 235 valid responses from a diverse range of respondents from various demographics and backgrounds. Among the respondents, 51.6\% identified as students, 42\% as working professionals, and 4.7\% as business owners. Regarding gender distribution, 77\% of respondents were male, 19.7\% were female, and 3.3\% preferred not to disclose their gender. 
Regarding the adoption timeline of UPI (figure \ref{fig:figures_row1}(a)), we can conclude that the diverse range of adoption dates reflects users' varying entry points into the UPI ecosystem, often aligning with significant events like demonetization and pandemic-related lockdowns.

\begin{figure}[h!]
  \centering
  \subfloat[]{\includegraphics[width=0.4\textwidth]{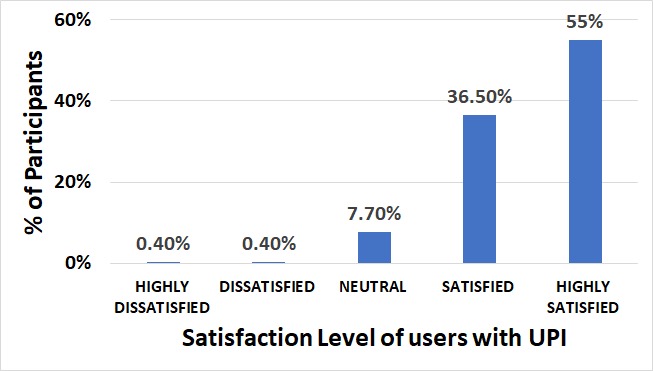}}\hspace{0.05\textwidth}
  \subfloat[]{\includegraphics[width=0.4\textwidth]{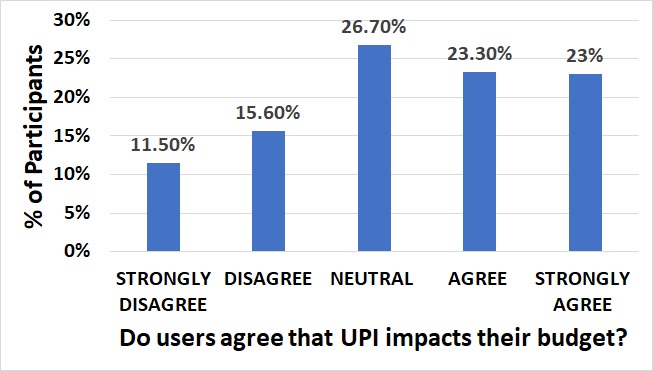}}\hspace{0.05\textwidth}
  \caption{(a) Satisfaction Level with UPI (b) UPI Impact on Budget  }
  \label{fig:figures_row2}
\end{figure}

To understand the impact of UPI adoption, 74.2\% of the survey respondents admitted that UPI had led to increased spending, while only 7\% said that it had decreased spending. Additionally, 18.8\% of the survey respondents felt no changes in their spending habits after adopting UPI. Regarding satisfaction levels (figure \ref{fig:figures_row2}(a)), 55\% of the survey respondents indicated they were highly satisfied with their overall UPI experience. Additionally, 36.5\% reported satisfaction, and 7.7\% remained neutral in their assessment. Therefore we can conclude that 91.5\% (55\% + 36.5\%) of the respondents are satisfied with their overall experience using UPI for payments. When asked to rate their convenience with UPI for making payments (figure \ref{fig:convenience_upi}), the survey results indicate that 69.50\% of respondents rated  "Highly convenient", 25.7\% rated "Convenient, and 3.7\% were "Neutral" on the same. Therefore we can conclude that 95.2\% (69.5\% + 25.7\%) of the respondents find using UPI for payments convenient. 
Notably, only 2-3 respondents reported dissatisfaction with UPI as a payment method, indicating that most were satisfied with their experience.


Respondents reported a varied daily spending range using UPI, from Rs 50 (0.60 USD) to 1 lakh (1,202.66 USD). On average, 15.7\% mentioned a daily spending of around Rs 200 (2.41 USD). Regarding the frequency of UPI usage for payments (figure \ref{fig:figures_row1}(b)), a significant 81.1\% use it daily, 15.6\% use it weekly, 1.1\% use it monthly, 1.90\% use it occasionally, and 0.30\% use it rarely. This data highlights the widespread and frequent adoption of UPI among surveyed individuals.

\begin{figure}[h!]
  \centering
  \subfloat[]{\includegraphics[width=0.4\textwidth]{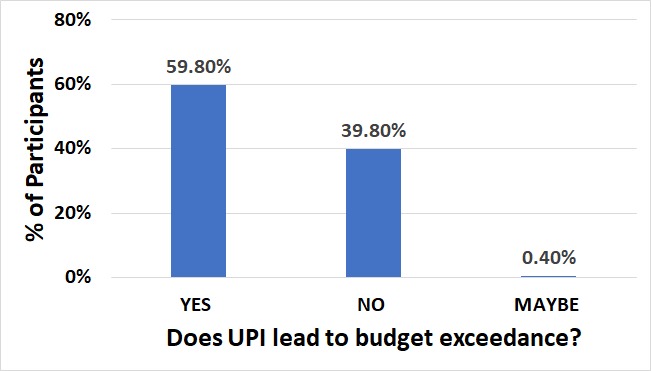}}\hspace{0.05\textwidth}
  \subfloat[]{\includegraphics[width=0.4\textwidth]{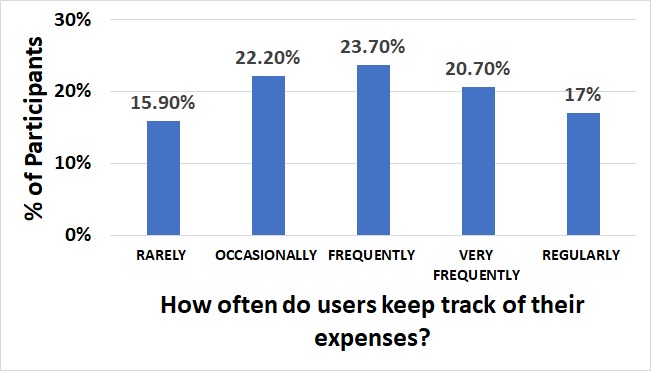}}\hspace{0.05\textwidth}
  \caption{(a) Budget exceedance due to UPI  (b) How often do people keep track of transaction history? }
  \label{fig:figures_row3}
\end{figure}

\begin{figure}[h!]
  \centering
  \includegraphics[width=0.4\textwidth]{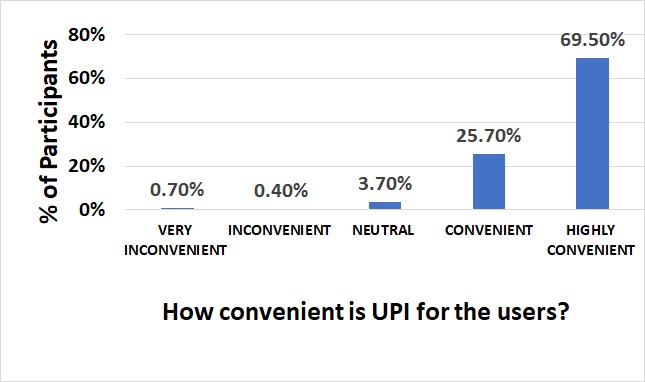}
  \caption{Convenience of making payments using UPI}
  \label{fig:convenience_upi}
\end{figure}

When asked about overspending due to UPI usage (figure \ref{fig:figures_row3}(a)), 59.8\% of respondents acknowledged experiencing such situations, while 39.8\% believed they had not. Additionally, respondents had diverse opinions when asked about the impact of using UPI on their budget adherence (figure \ref{fig:figures_row2}(b)); approximately 23\% strongly agreed, 23.3\% agreed, 26.7\% remained neutral, 15.6\% disagreed, and 11.5\% reported strongly disagree, indicating that UPI has had an impact on their ability to stick to their budget and control overspending.

Respondents employed various methods to manage their expenses, like checking their transaction history, Splitwise \cite{spl} SBI YONO \cite{yono}, Samsung Notes \cite{snotes}, using excel sheets and dedicated finance apps like MyMoneyy \cite{mymoney}, Honeybook \cite{honeybook}, Monefy \cite{my}, Payhawk \cite{payhawk} and Money Manager \cite{mm}. 
This observation highlights diverse financial management approaches among UPI users, showcasing varied responses and opinions. A significant proportion acknowledges increased spending along with high overall satisfaction with UPI and its convenience. Our comprehensive analysis of various financial tracking and budget management apps in the Indian market is summarized in Table \ref{tab:competitive_analysis} .


\subsection{Graphs for Quantitative Results}

\subsection{ Qualitative Analysis (Analysis of Interview Results)
}




\subsubsection{\textbf{Exploring UPI Usage for Transactions: Insights and Perspectives}}\hfill\\
Respondents appreciated UPI's convenience and cashless transactions, obviating the requirement to carry tangible currency. They consistently praised the system's remarkable speed and efficiency in handling small transactions.





According to a State Bank of India (SBI) report reveals that with every Rs 1 (0.012 USD) increase in UPI transactions, there is an 18 paisa (0.0022 USD) decline in debit card transactions, leading to a decrease in ATM visits from 16 to 8 times a year [44]. Additionally, UPI transactions create a digital record that helps users maintain a record of transactions and financial documentation, collectively contributing to high satisfaction levels among UPI users.


\emph{\textbf{“I love UPI! It's a fantastic way to spend money without dealing with change or carrying cash. You can pay just Rs 1 (0.012 USD) to a whopping 1 lakh rupees (12,028.79 USD) in one go. It's one of the best things in this progressive era.”-[P1]}}

Certain respondents, like a lawyer and clinic owner, emphasized UPI's role in expediting payments and enhancing business efficiency and financial processes. Notably, UPI's absence of processing fees for peer-to-peer transactions stands out, benefiting merchants and customers and distinguishing it from credit and debit cards.



\emph{\textbf{"I encourage my customers to use UPI or cash for payments instead of credit or debit cards to avoid the 2-3\% transaction fees, which either come from my earnings or are passed on to the customer, potentially harming my reputation. UPI offers a real-time, no transaction fees and transparent payment solution."-[P2]}}

The rise of UPI adoption brings risks, with a surge in scams and fraud. Respondents shared phishing incidents, where malicious actors posed as known contacts or UPI platforms to trick users into sharing sensitive details raising concerns about user vulnerability to cyber threats. Notably, in 2022-23, over 95,000 fraud cases related to UPI transactions were recorded, a notable increase from 84,000 incidents in 2021-22 \cite{singh}.To address these issues, users can take protective steps like avoiding sharing sensitive information like OTPs and bank details, updating UPI apps, enabling two-factor authentication, and being cautious with unsolicited payment requests.

\emph{\textbf{“I once got a message asking me to send money, and they wanted my bank OTP. Sadly, I gave it to them, and things got messed up. They took a big chunk of cash, around Rs 10,000, from my bank account. Fortunately, my father worked at a bank and helped fix it. I learned a harsh lesson about being careful with UPI and not sharing important info.”-[P3]}}



\subsubsection{\textbf{Factors Influencing Spending Behavior}}\hfill\\
A few respondents strongly preferred using UPI as a payment option as it simplifies transactions, reduces dependency on cash, and encourages frequent expenditure. This shift contributes to a more efficient, cashless economy, revolutionizing how people handle their finances and everyday purchases. 


\emph{\textbf{"I spend more often with UPI because you pay directly from your bank account. UPI combined with the rise of Swiggy or Zomato or Zepto or Blinkit, etc., ordering online has become much more appealing and easy. Without UPI, I might not have been spending much on food online. Sometimes, I regret purchases because you only needs a phone with Wifi/mobile data"-[P4 \& P5]}}

This digital payment method has reshaped how individuals perceive and manage their finances, sometimes leading to more frequent and impulsive spending patterns.
Many individuals struggle with financial regret from overspending, the need for expense mindfulness, and less-than-ideal financial choices. The root cause lies in the stark contrast between physical cash and the abstract nature of digital transactions like UPI payments. Physical cash provides a tangible connection, whereas digital transactions involve mere numbers on a screen, leading to spending without much thought and subsequent feelings of regret and guilt.


\emph{\textbf{“Digital money does not give you guilt when spending. When you spend paper money, you feel guilty. Cash feels tangible, like giving away something valuable when we spend it. But with UPI, it's just a number on the screen, and it doesn't feel the same way."-[P6 \& P7]}}

A diverse range of insights emerged when we asked users about their spending habits and where they generally used UPI. It was evident that UPI has found its way into various aspects of daily life, transforming how people manage their expenses. Users reported using UPI for essentials like rent, groceries, and bills, as well as for work-related expenses, leisure activities such as dining out and entertainment, education, online shopping, and even small everyday indulgences like street food and auto-rickshaw fares. 


UPI's integration into numerous applications has positioned it as a key player in personal and professional finances, particularly in critical business transactions. Its omnipresence showcases its transformative impact on the Indian economic landscape. Respondents acknowledge UPI's significant influence on spending behavior, budgeting habits, and financial discipline. The study identifies distinct and common patterns in UPI adoption among various user groups, emphasizing its broad utility in routine transactions and reshaping spending habits.

\subsubsection{\textbf{Attitude Towards Financial Management}}\hfill\\
Respondents highlight the significance of financial proficiency in UPI usage, stressing the need for personal finance education and expenditure restraint. Attitudes toward financial discipline varied, reflecting different comfort levels in managing finances using UPI, indicating individual variability in its impact on personal finances.

\emph{\textbf{"Personally, there are occasions when I exceed my initial spending plans, but I rely on finance tracking applications to ensure I stay within my budget.”-[P8]}}

\subsubsection{\textbf{Recommendations for Finance Tracking Applications}}\hfill\\
Users expressed the need for visualization tools like graphical representations and pie charts to help them understand their spending patterns. The ability to track expenses over time and categorize them was also a common request. Users want insights into where their money is going, allowing them to make more informed financial decisions.

\emph{\textbf{“I'd like my UPI app to allow me to restrict spending in certain categories, such as food, by using QR codes to track locations and deny transactions in those places. I want the app to generate a pie chart showing the average payments I make in different categories over a specific period.”-[P9]}}


Responses from various professional backgrounds revealed that the specific needs of their professions often shape users' expectations of financial management apps. 
Business professionals sought tools to manage business-related finances efficiently, while doctors aimed to collect patient payments and related transactions effectively. These insights underscore the importance of tailoring financial apps to meet the unique requirements of diverse professions, ensuring effective financial management across various occupational backgrounds.

\emph{\textbf{“As a doctor, it would be beneficial if I could easily keep track of all the patients who have paid their fees. Checking transaction histories repeatedly can be quite a hassle. So, having an automatically generated list of patients who have made payments, with a clear categorization of whether the payment has been completed, would be incredibly convenient.”-[P10]}}

Some respondents went beyond traditional budgeting and desired financial education and guidance within the app, along with receiving notifications and alerts when approaching spending limits or exceeding budgeted amounts. 


\emph{\textbf{“Besides handling expenses, it would be great if the app also offers guidance on achieving financial stability and discipline by providing tips and educational content to help users make better financial decisions.”-[P11]}}

\section{Insights from Phase 1 Evaluation}\label{sec:Requirements_ for_next_gen.tex}

Users expressed a desire to incorporate customizable spending limits and real-time notifications into the redesigned app, ensuring users stay aligned with their financial goals. In addition, users frequently requested better ways to visualize their spending patterns with the integration of detailed graphical tools, including pie charts, providing users with a clear representation of their financial habits over time. Financial education also emerged as a significant gap in many apps, with users seeking educational pop-ups and financial tips with a built-in “Financial Education” component which offers users practical financial quotes and suggestions through timely pop-ups during transactions, helping them stay informed on budgeting, saving, and responsible spending. By delivering these insights in real-time, the app can enhance users’ financial literacy and promote better decision-making. Lastly, our results highlighted a strong user preference for automated expense tracking that balances privacy concerns associated with accessing the financial details of the users. 

Our comprehensive analysis of various financial tracking and budget management apps in the Indian market is summarized in Table \ref{tab:competitive_analysis}. These applications offer diverse features tailored to meet individual financial requirements, ranging from automated organization of transactional data to robust expense-tracking capabilities. Users can leverage these tools to elevate their financial management skills. Specific features like detailed graphical representations of expenditure patterns and spending thresholds with reminders are unevenly available across platforms. 

We examined which features already exist in these apps, particularly those related to budget tracking, spending categorization, and financial planning. A key observation was that although many of these apps excel in budget tracking and expense management, none of them support UPI payments. This presents a significant limitation, as users are forced to switch between multiple apps to manage expenses and make payments. In conclusion, while existing financial tracking apps offer valuable features, there is room for improvement in aligning with the recommendations outlined. Future developments should focus on bridging these gaps to provide users with a more holistic and empowering financial management experience.

  




\begin{table*}
    \centering
    \tiny
    \begin{tabular}{|p{1cm}|p{3cm}|p{1cm}|p{3cm}|p{3cm}|p{1cm}|}
         \hline
         & FEATURES & YEARLY SUBSCRIPTION COST & PROS & CONS & SUPPORTING PLATFORMS \\
         \hline
        FinArt \cite{fin} & Automatically organizes transactional SMS into expenses, budget, and bill reminders, even when the app is closed. & Rs 799  (9.61 USD) & Data stored on Drive, Auto Tracking, spend analysis dashboards, and budget setting. Separates personal and business accounts for comprehensive financial management. & Setup complexity,  ads persist after subscription, and occasional difficulty in SMS tracking. & ANDROID, PC, IOS \\
         \hline
        Money Manager  \cite{mm} & The app efficiently handles dual accounts, ensuring separate tracking of income and expenses. It offers offline/cloud backup and Data Export functionalities. & Free & Diverse payment method profiles, encompassing cash and cards, support competitive analysis. Easily create new profiles based on specific needs. & The application takes a long time if currencies are changed to optimize expenses and income. Most fees are non-editable.  & ANDROID, IOS \\
         \hline
        Monefy  \cite{my} & The app is straightforward; the infographics make it vibrant and easy to use. & Rs 180 (2.17 USD)  & Set bill reminders, manage multiple accounts and easily import data through Excel. & Frequent prompts for premium upgrades  & ANDROID, PC ,IOS \\
         \hline
        Wallet  \cite{wall}& Users can use it for different currency rates worldwide. The Group Sharing option is also available in the application & Rs 700 (8.42 USD) & Monitor group expenses in the totals tab, track recent costs in the activity tab, and official API support available.  & Accounts auto-disconnect; unable to modify the initial login currency.  & ANDROID, IOS \\
         \hline
        Splitwise \cite{spl} & App simplifies bill splitting , not for personal expense tracking. Form groups, manage shared bills, and track dues effortlessly. & Rs 499 (6 USD)   & Clean UI. Manages multiple wallets, links bank accounts or Crypto wallets directly, and tracks transactions seamlessly (premium feature). & Tracking and analyzing personal finances can be challenging.  & ANDROID, PC, IOS  \\
         \hline
        Spendee \cite{spen}  & The app provides user-friendly expense and income tracking with easy-to-use tags for sorting costs efficiently. & Rs 900 (10.43 USD)  & Effortlessly track expenses, share budgets, view spending analytics, and easily monitor debt progress.  & Certain features are restricted behind a paywall. & ANDROID, PC, IOS \\
         \hline
        GoodBudget \cite{gb} & Budget management application. The app lets you create and refill multiple budgets (envelopes) at intervals like monthly or weekly.  & Rs 5817 (70 USD)  & Create multiple wallets, set up recurring payments, track credit card and debt. Link bank accounts and enable group sharing (premium only). & Simple UI. Adding transactions is tedious, and there's no API support.  & ANDROID, PC, IOS \\
         \hline
    \end{tabular}
    \caption{\textbf{Comparison of Personal Finance Tracking Apps}}
    \label{tab:competitive_analysis}
\end{table*}

Using the insights from \textbf{Phase 1} evaluations and comparative analysis of existing apps, we identified a list of features to be integrated into the redesign prototype of the UPI mobile application. The next section presents all the relevant details associated with prototype design and corresponding usability testing.

\section{Phase 2: Initial Prototype}\label{sec:phase-2-initial-prototype}

The initial prototype was created using Figma \cite{figma}. The process involved brainstorming and merging ideas from \textbf{Phase 1}, ensuring the implemented features were well-planned and aligned with user preferences. Additionally, we analyzed existing financial tracking apps to identify already implemented features and gaps, which informed our approach to feature inclusion and implementation. This iterative development and testing process has provided valuable insights into how these enhancements might influence spending behaviors for UPI users .

Due to the stringent regulations imposed by the Reserve Bank of India (RBI) and the Government of India on financial apps, creating a full-fledged UPI payment app was not feasible. Thus, we chose PhonePe \cite{phonepe}, India's most popular UPI app, as our prototype’s foundation. This choice allowed us to efficiently incorporate innovative features and suggestions while leveraging an established platform for user testing. Our research aimed to enhance the user experience by addressing users' financial management needs within a familiar and widely-used app.

\subsection{Features}\label{sec:features}  

\subsubsection{\textbf{Expense Tracker}}
\begin{figure*}
    \centering
    \includegraphics[width=1\textwidth]{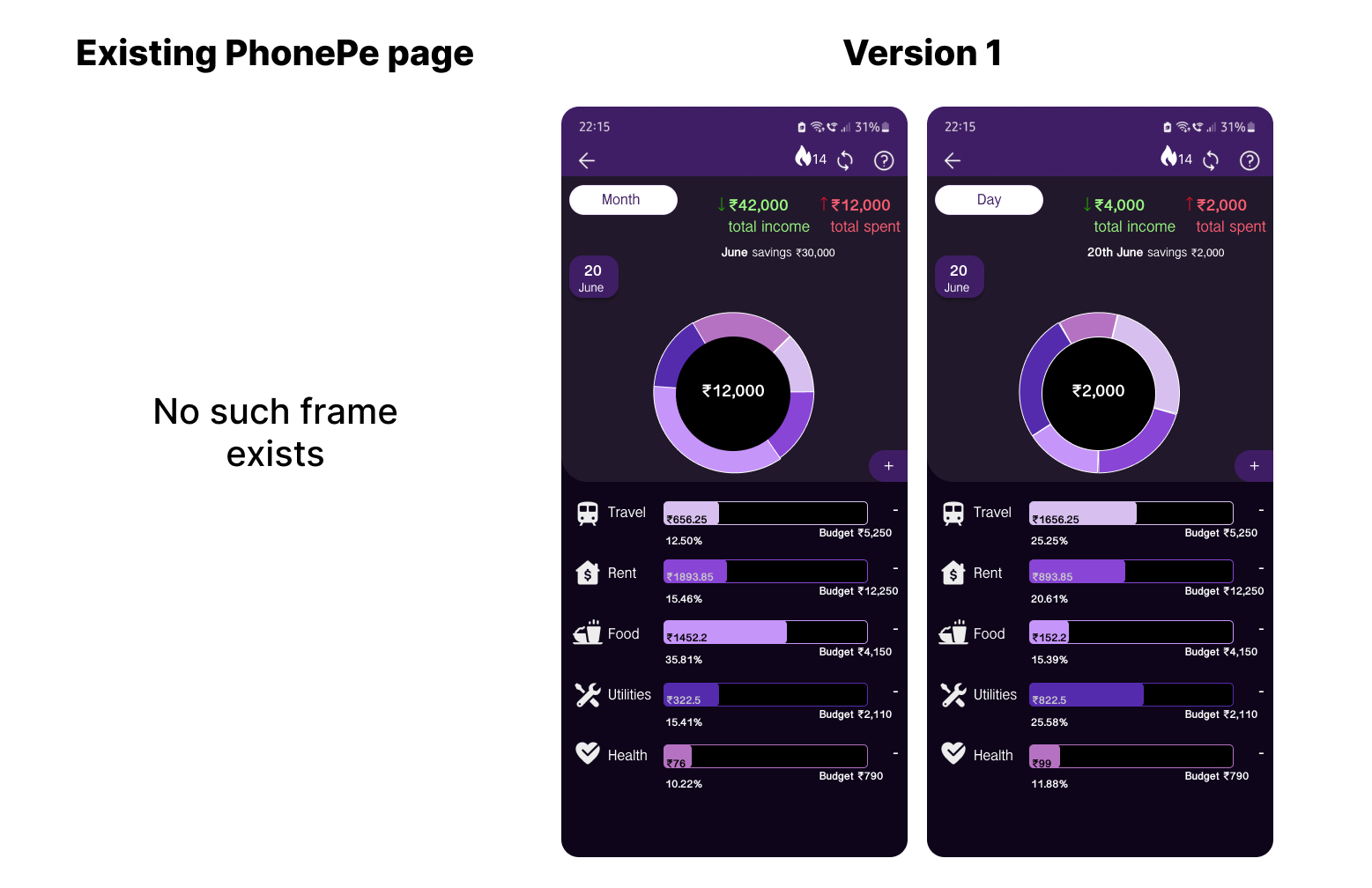}
    \caption{\textbf{Expense Tracker }}
    \label{fig:1v1}
\end{figure*}

\hfill\\
The redesigned PhonePe prototype includes an Expense Tracker as shown in the Figure \ref{fig:1v1} that enables users to track their spending across categories like food, travel, and more. This feature provides detailed day-wise and month-wise summaries, displaying total incoming and outgoing cash flow, budget, and spending patterns. It also indicates whether users exceeded their budgets. The tracker provides an overall financial summary, including income, expenses, and savings, while using color-coding—red for overspending and green for savings—allowing users to quickly assess their financial status and adjust their spending habits accordingly.

\subsubsection{\textbf{Category Management}}
\begin{figure*}
    \centering
    \includegraphics[width=1\textwidth]{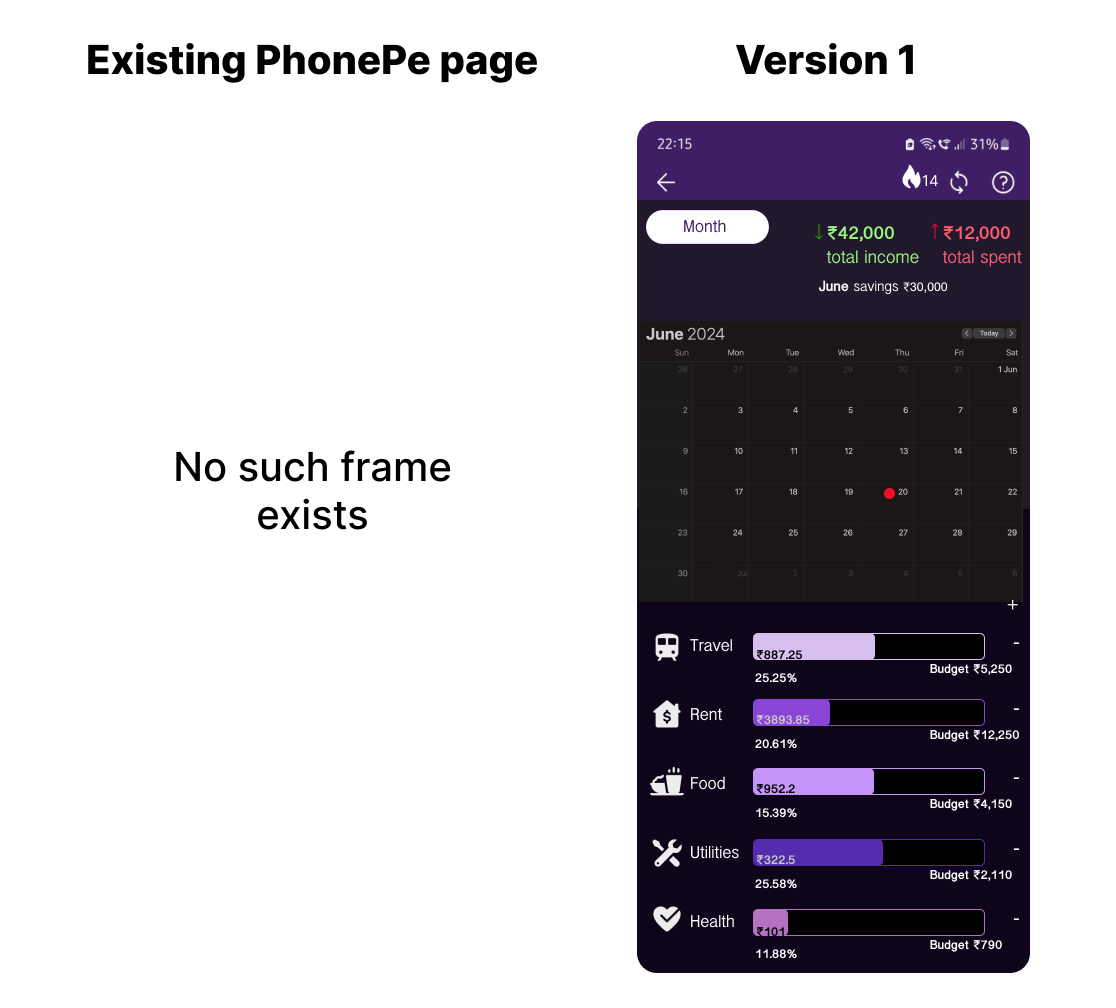}
    
    \caption{\textbf{Categorization of Expenses} }
    \label{fig:2v1}
\end{figure*}

\hfill\\
Users can personalize their experience by creating new categories or removing irrelevant ones  through the use of "+" and "-" icons as shown in Figure \ref{fig:2v1}. This Category Management feature ensures that users can organize their spending according to their needs, making it easier to track expenses that are most relevant to them.

\subsubsection{\textbf{Payment Selection}}
\begin{figure*}
    \centering
    \includegraphics[width=1\textwidth]{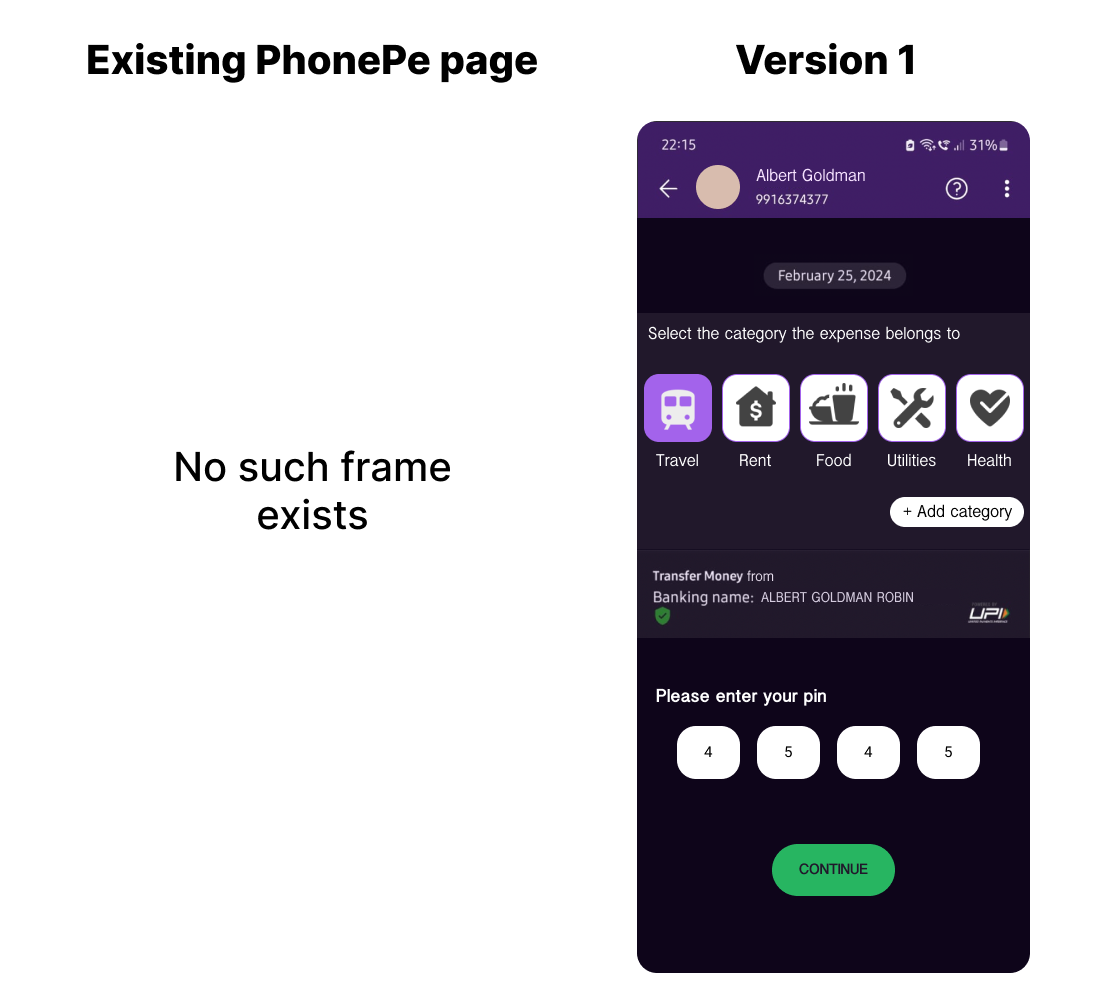}
    \caption{\textbf{Payment Type Selection} }
    \label{fig:3v1}
\end{figure*}

\hfill\\
The prototype also features a Payment Selection option as shown in Figure \ref{fig:3v1} , enabling users to categorize their expenses before making payments. This categorization is automatically integrated into the expense tracker, providing users with clear insights into areas where they are overspending and where adjustments might be necessary for better financial management.

\subsubsection{\textbf{Balance Display}}
\begin{figure*}
    \centering
    \includegraphics[width=1\textwidth]{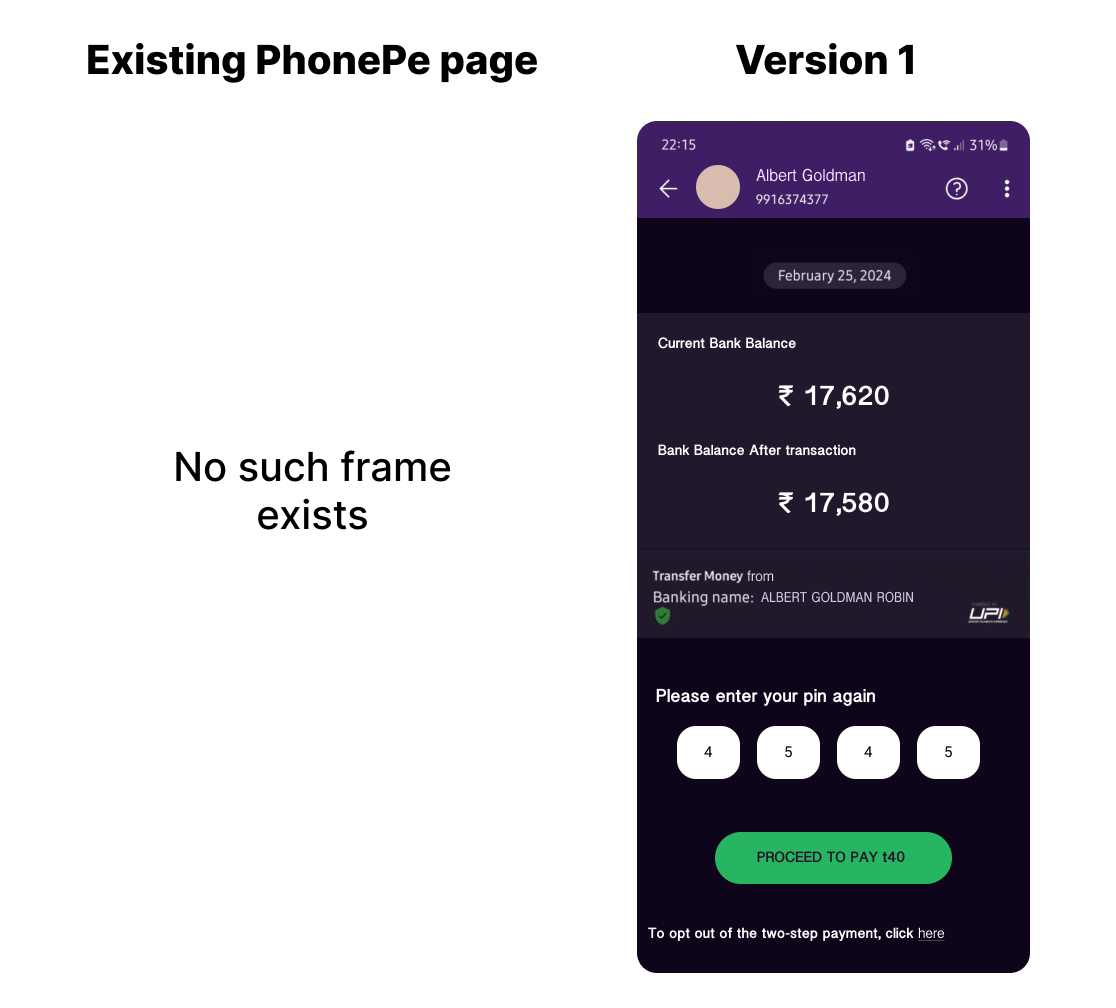}
    \caption{\textbf{Balance Display}}
    \label{fig:4v1}
\end{figure*}

\hfill\\
A key component of the prototype is the Balance Display feature in Figure \ref{fig:4v1} , which shows users their current balance before making a payment and the balance remaining after the transaction. This serves as a double-check mechanism, helping users make informed spending decisions and plan for future expenses by providing real-time visibility into their financial status.

\subsubsection{\textbf{Financial Education Pop-Ups}}
\begin{figure*}
    \centering
    \includegraphics[width=1\textwidth]{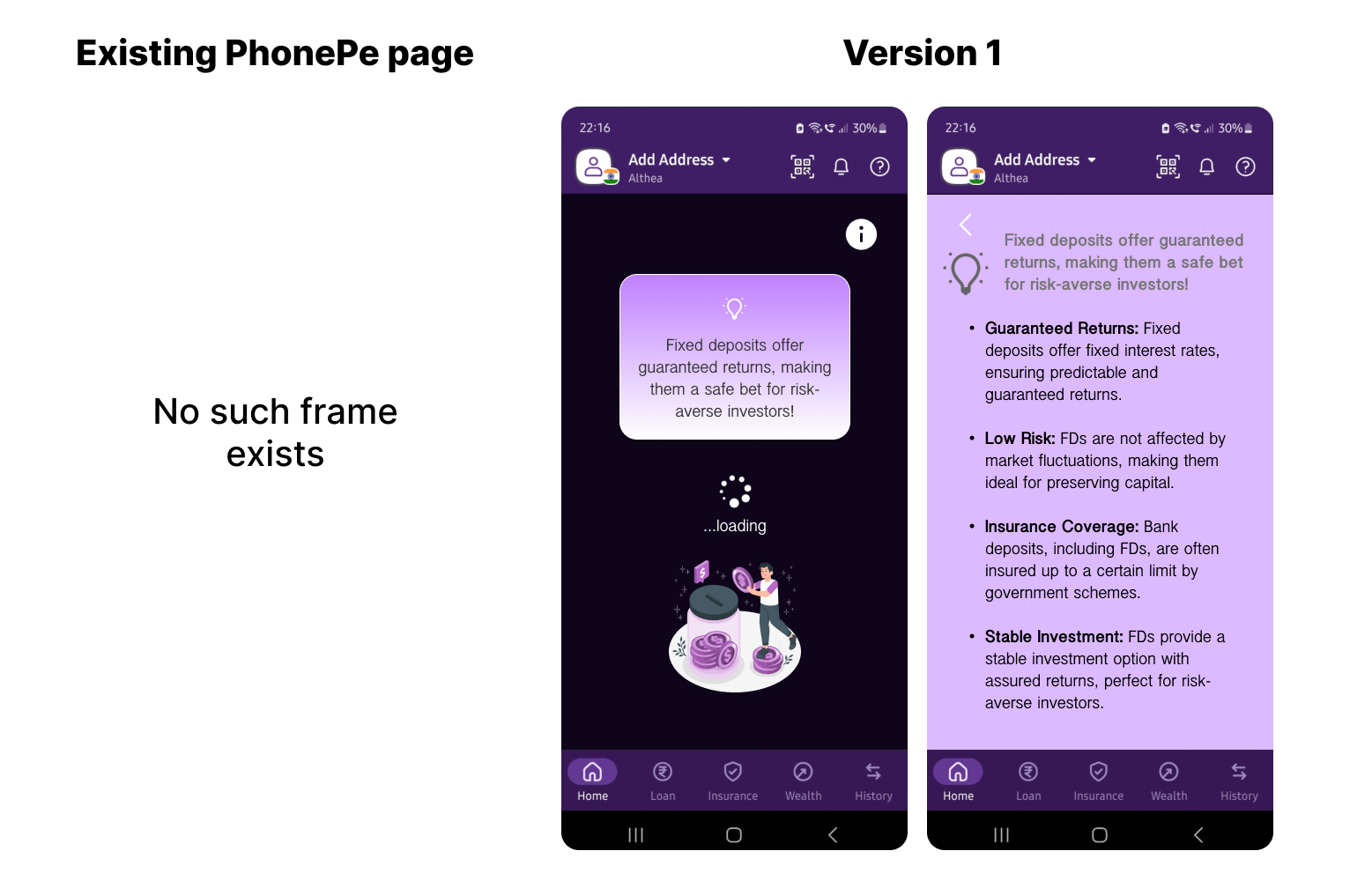}
    \caption{\textbf{Financial Education Pop-Ups} }
    \label{fig:5v1}
\end{figure*}

\hfill\\
To promote financial literacy, as shown in Figure \ref{fig:5v1} , the prototype incorporates a Financial Education Pop-Up feature, offering users helpful financial quotes and tips through interactive pop-ups. These pop-ups can be clicked for more detailed information, enhancing user engagement and making financial education an integral part of their everyday transactions.

\subsubsection{\textbf{Budgeting Limitations}}
\begin{figure*}
    \centering
    \includegraphics[width=1\linewidth]{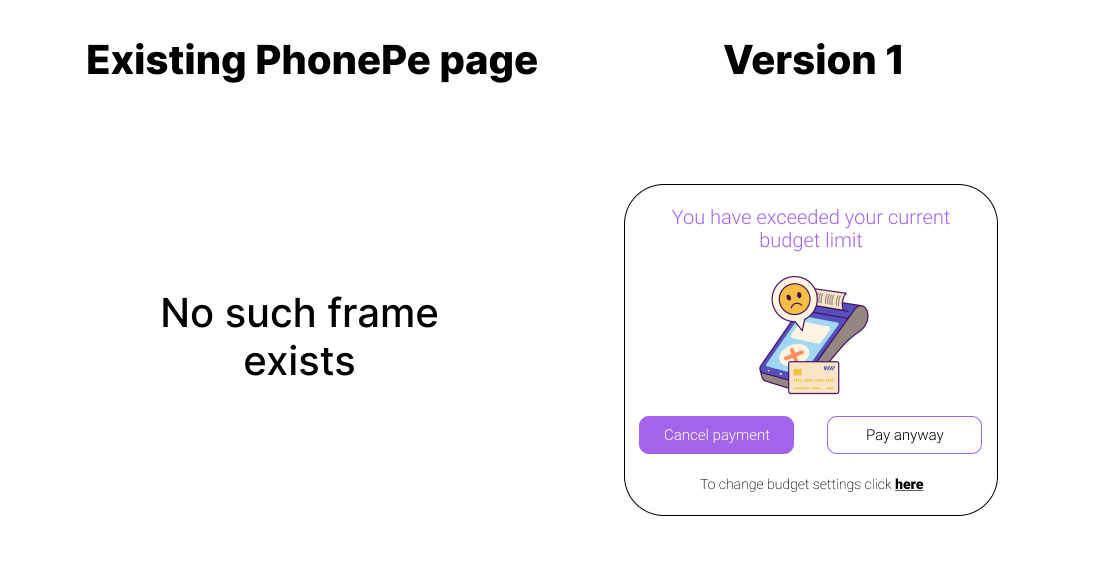}
    \caption{\textbf{Budegting Limitations}}
    \label{fig:6v1}
\end{figure*}

\hfill\\
The Budgeting Limitations feature empowers users to set overall budget limits within the app as shown in Figure \ref{fig:6v1}, which they can easily toggle on or off as needed. This functionality helps users maintain control over their spending by establishing personal financial boundaries, making it easier to manage their monthly finances and prevent overspending.

\subsubsection{\textbf{Saving Streak Feature}}
\begin{figure*}
    \centering
    \includegraphics[width=1\linewidth]{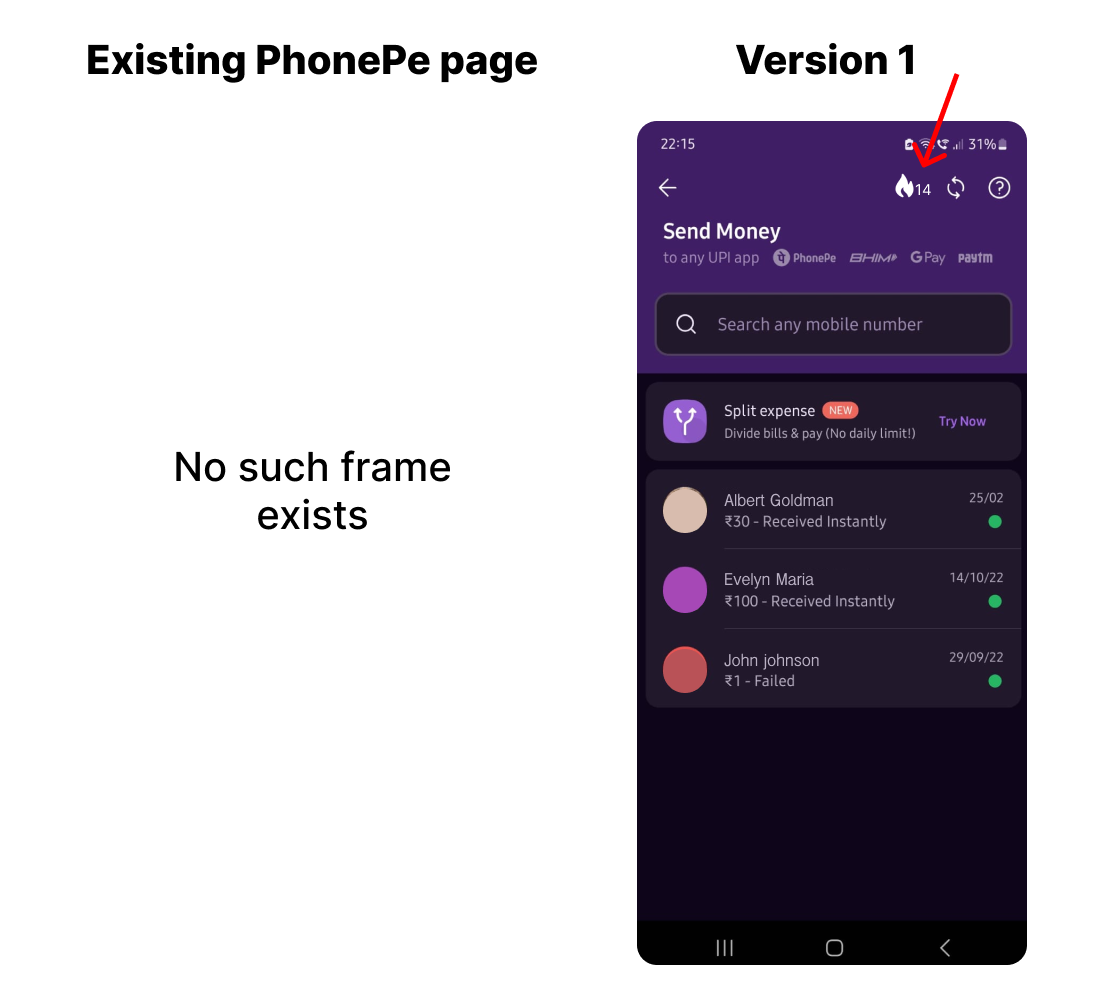}
    \caption{\textbf{Saving Streak}}
    \label{fig:7v1}
\end{figure*}

\hfill\\
A gamified element introduced in the prototype is the Saving Streak Feature as shown in  Figure \ref{fig:7v1} , which displays users' savings streaks at the top of the home screen. This feature highlights how long a user has been maintaining their budget, encouraging consistent financial habits by gamifying the experience of saving, making it both a fun and rewarding activity.

\subsubsection{\textbf{Spam Detection}}
\begin{figure*}
    \centering
    \includegraphics[width=1\linewidth]{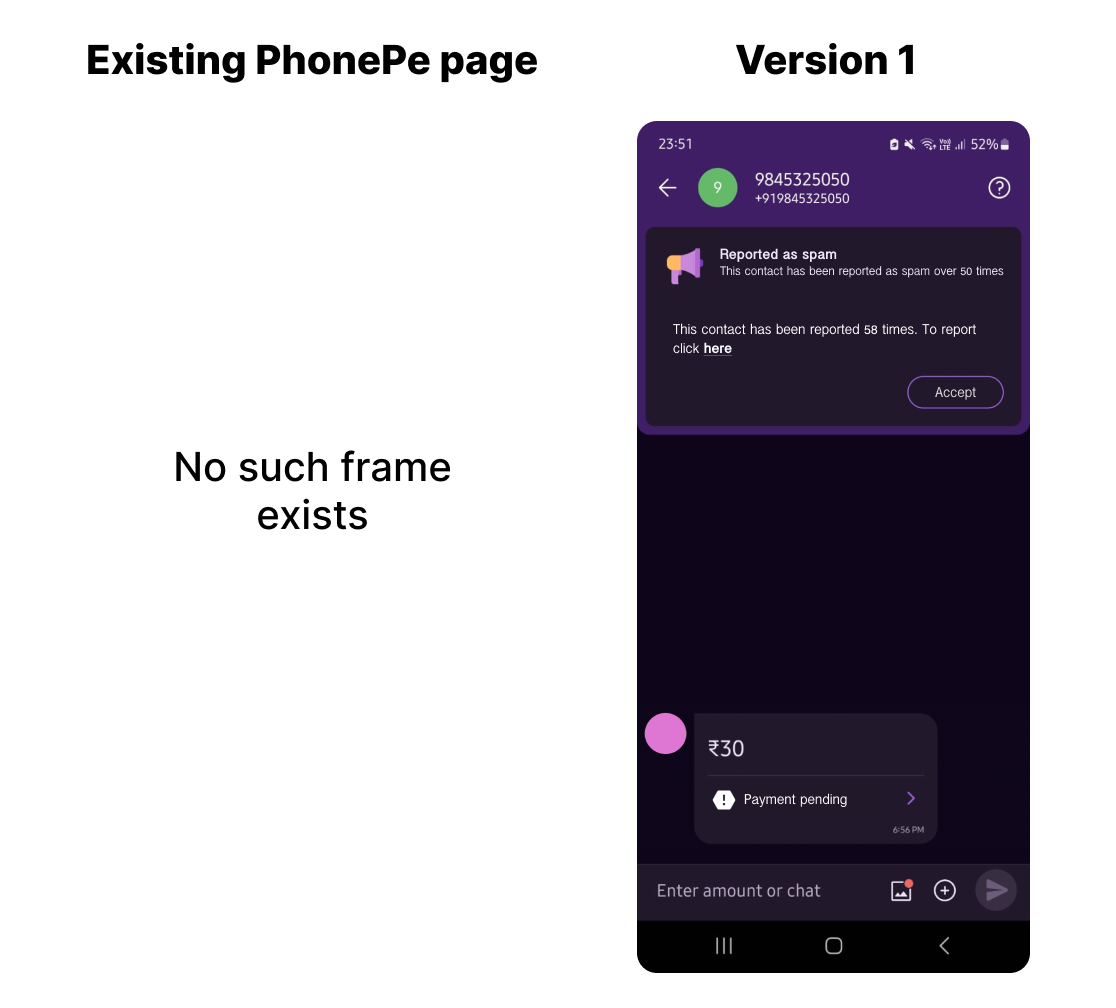}
    \caption{\textbf{Spam Detection}}
    \label{fig:8v1}
\end{figure*}

\hfill\\
Due to increasing scams related to UPI as mentioned in \S\ref{sec:related_work}, we implemented a feature of “spam detection” as shown in the Figure \ref{fig:8v1} . If a UPI ID is being reported as scam by multiple people, it would most probably be a scam too and hence alerting the user that they need to get cautious when they are making payment to this particular id (just like how Truecaller spam detection works \cite{truecaller}).

\subsubsection{\textbf{Delay Feature for Large Payments}}
\begin{figure*}
    \centering
    \includegraphics[width=1\linewidth]{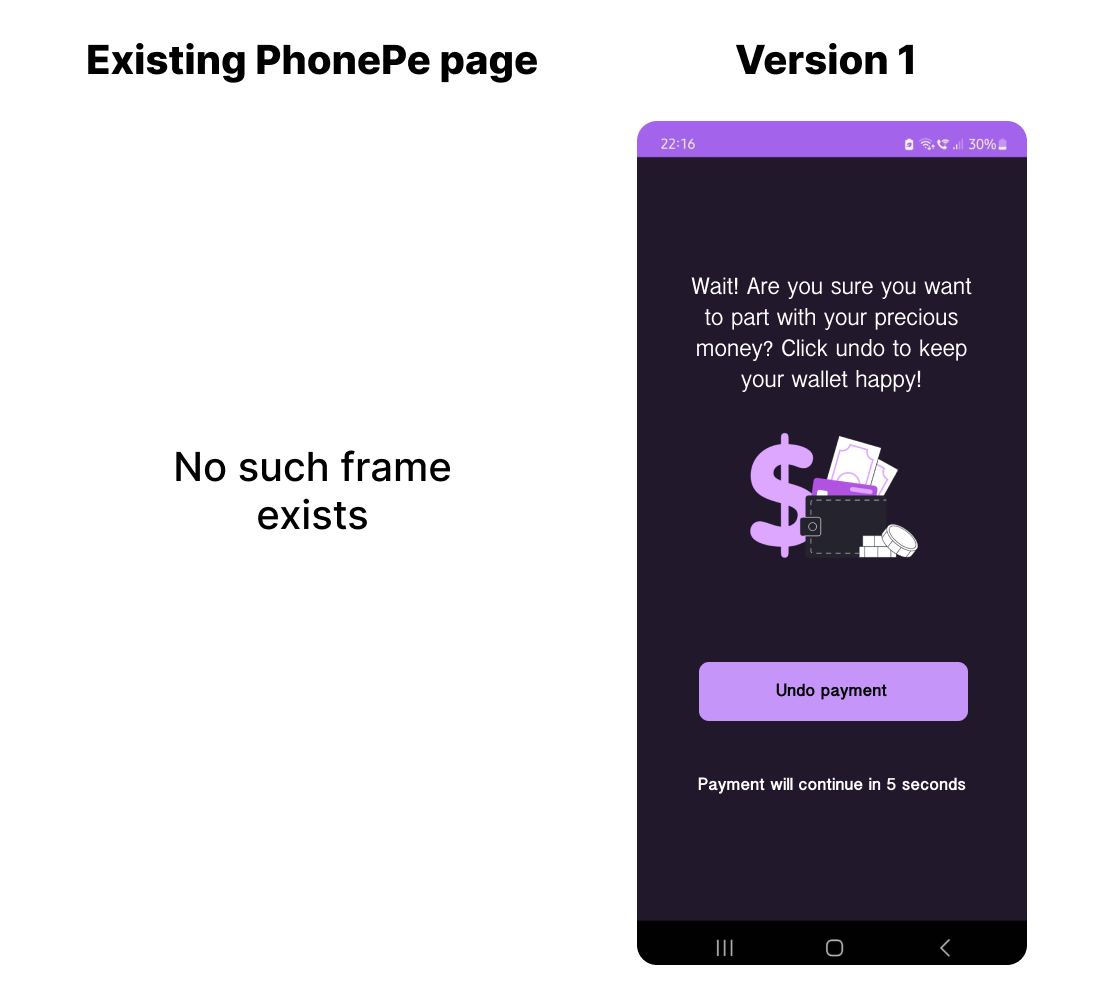}
    \caption{\textbf{Delay Feature}}
    \label{fig:9v1}
\end{figure*}

\hfill\\
The prototype includes a Delay Feature for Large Payments as shown in the Figure \ref{fig:9v1} . When users initiate a significant payment, a timer is triggered, allowing them to reconsider the transaction briefly. This feature simulates the "natural hesitation" seen in physical transactions, providing users with a moment to be more mindful of their spending before confirming larger expenses.

\section{Phase 2: Methodology for Usability Testing }\label{sec:phase-2-methodology}

Following the feedback received in Phase 1, we developed an enhanced prototype of the PhonePe app (version 2), incorporating user suggestions and refining key features based on their needs. To evaluate this updated prototype, we conducted a moderated usability study, which provided valuable insights into how users interacted with and responded to the newly implemented features. This study helped us assess user satisfaction and identify areas for improvement, ensuring the redesigned PhonePe app prototype better met user expectations and functionality requirements.

\subsection{Study Design and Data Collection}
The study followed a structured, multi-stage approach:

\noindent\textbf{Stage 1: Recruitment:}
We advertised our research study across various online platforms, inviting interested users to sign up. A total of 34 users, representing a diverse range of ages, educational backgrounds, and technological proficiency levels, responded (see Table \ref{table:demographic}). All 34 users were invited to participate in the study. Before beginning, each participant was provided with a detailed overview of the study’s objectives and an overview of the prototype’s functionalities.

\noindent\textbf{Stage 2: Prototype Walkthrough:}
Participants were given access to the developed prototype via Figma \cite{figma}. They were also provided with a privately shared YouTube video that demonstrated the prototype’s features in detail. During this stage, participants were encouraged to ask questions to ensure they fully understood the prototype's functionalities.

\noindent\textbf{Stage 3: Feedback Questionnaire:}
After exploring the prototype, participants were asked to complete a feedback questionnaire via Google Forms. The questionnaire was designed using the \textbf{Technology Acceptance Model (TAM)} \cite{tam1989}, focusing on two core factors: \textbf{Perceived Usefulness (PU)} and \textbf{Perceived Ease of Use (PEOU)}. This model helped assess user attitudes towards the new features, such as the expense tracker and budgeting functionalities, and provided a structured approach to understanding factors influencing user acceptance of the redesigned PhonePe app. The detailed questionnaire used in this phase is available in Appendix \ref{appendix:phase_2_questionnaire}.

\noindent\textbf{Stage 4: Follow-Up Interviews:}
To gain deeper insights, we conducted follow-up interviews with 20 participants (13 male, 7 female) who expressed interest in further discussing their experiences. Each participant gave explicit consent for recording their responses. The interview questions were based on the feedback questionnaire itself. These interviews offered rich qualitative data, providing a more richer understanding of the participants’ experiences. All interviews were transcribed for further analysis.

\subsection{Data Analysis}
The data collected in \textbf{Phase 2} was analyzed using both quantitative and qualitative approaches to gain a comprehensive understanding of user feedback and experiences.

For the \textbf{quantitative analysis}, we aggregated the responses from the feedback questionnaire to identify trends and patterns related to the perceived usefulness (PU) and perceived ease of use (PEOU) of the prototype's new features, such as the expense tracker and budgeting functionalities. Key metrics, such as the percentage of users who found the features beneficial or easy to use, were calculated and presented as aggregate numbers to provide a clear overview of user acceptance and satisfaction.

For the \textbf{qualitative analysis}, we performed a thematic analysis of the interview transcripts using an inductive approach. This method allowed us to identify recurring themes and patterns that emerged from participants' responses, providing deeper insights into their experiences and perceptions. Themes such as user engagement, suggestions for improvements, and the impact of the new features on financial management were coded and categorized. Thematic analysis helped uncover richer perspectives that complemented the quantitative findings, providing a richer understanding of user interactions with the prototype.


\subsection{Ethical Considerations}
In this phase also, we ensured that we followed all the ethical practices as already mentioned in \S\ref{sec:p1ethical}.

\subsection{Limitations}
In addition to our limitations mentioned in \S\ref{sec:p1limit}, due to regulatory constraints imposed by the Reserve Bank of India and the Government of India, developing an actual financial payment app is not feasible. Regulatory frameworks such as the 'Master Directions on Digital Payment Security Controls' and the 'Regulation of Payment and Settlement Systems Act, 2007' outline stringent requirements that limit our ability to create a fully functional financial application. Although we conducted usability testing on the redesigned UPI app prototype and incorporated the user feedback to develop the final prototype, it would be interesting to conduct an additional usability testing of the final prototype to further refine the developed prototype. This is part of our future work.

\begin{table}[t]
	\vspace{-1em}
    
	\begin{tabular}{|p{2cm}|p{8cm}|} 
		\hline
		Gender  & Female : 15, Male : 19\\
		\hline
		Age& Min : 20, Max : 54 , Avg :   40.15, SD : 15.71\\ 
		\hline
        Occupation &  Working Professionals : 10, Student : 19,  Business :  4, Home Maker : 1\\ 
		\hline
	  
	\end{tabular}
    \vspace{1em}
	\caption{\textbf{
Participants Demographics for the Usability Testing (Phase 2)}}
	\label{table:demographic}
\end{table}

\section{ Phase 2: Results of Usability Testing}\label{sec:results-usability-testing}
\subsection{Quantitative Analysis}

In Phase 2, we received 34 responses from a diverse range of respondents across various demographics and backgrounds. Among the respondents, 64.7\% were aged between 18 and 30, 23.5\% between 30 and 45, and 11.8\% were aged 45 and above. Regarding occupation, 58.8\% were students, 26.5\% were working professionals, 8.8\% were business owners, and 2.9\% were homemakers. Regarding gender distribution, 57.6\% of the respondents were male, while 42.4\% were female. 

\begin{figure}[h!]
  \centering
  \subfloat[]{\includegraphics[width=0.4\textwidth]{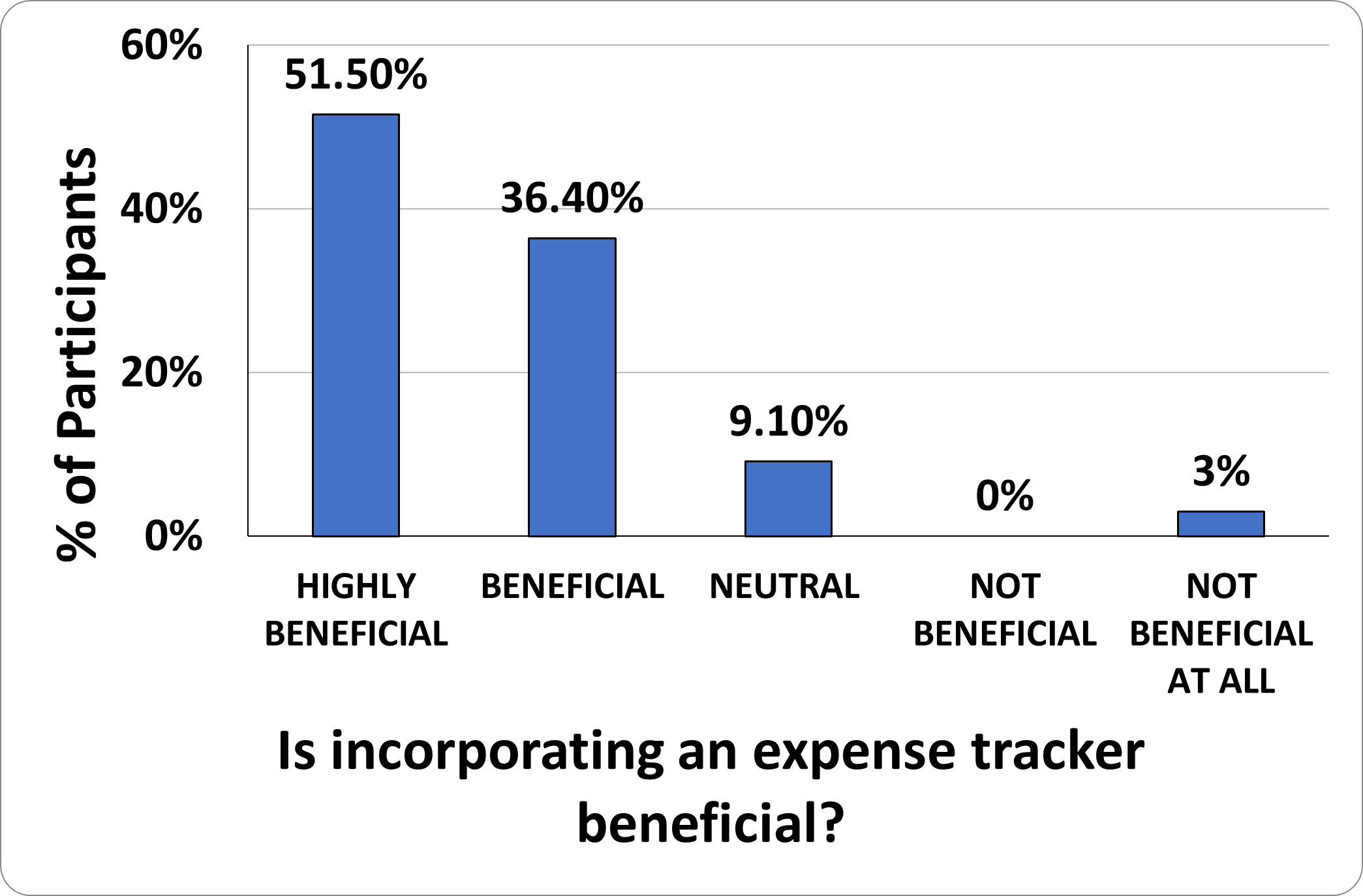}}\hspace{0.05\textwidth}
  \subfloat[]{\includegraphics[width=0.4\textwidth]{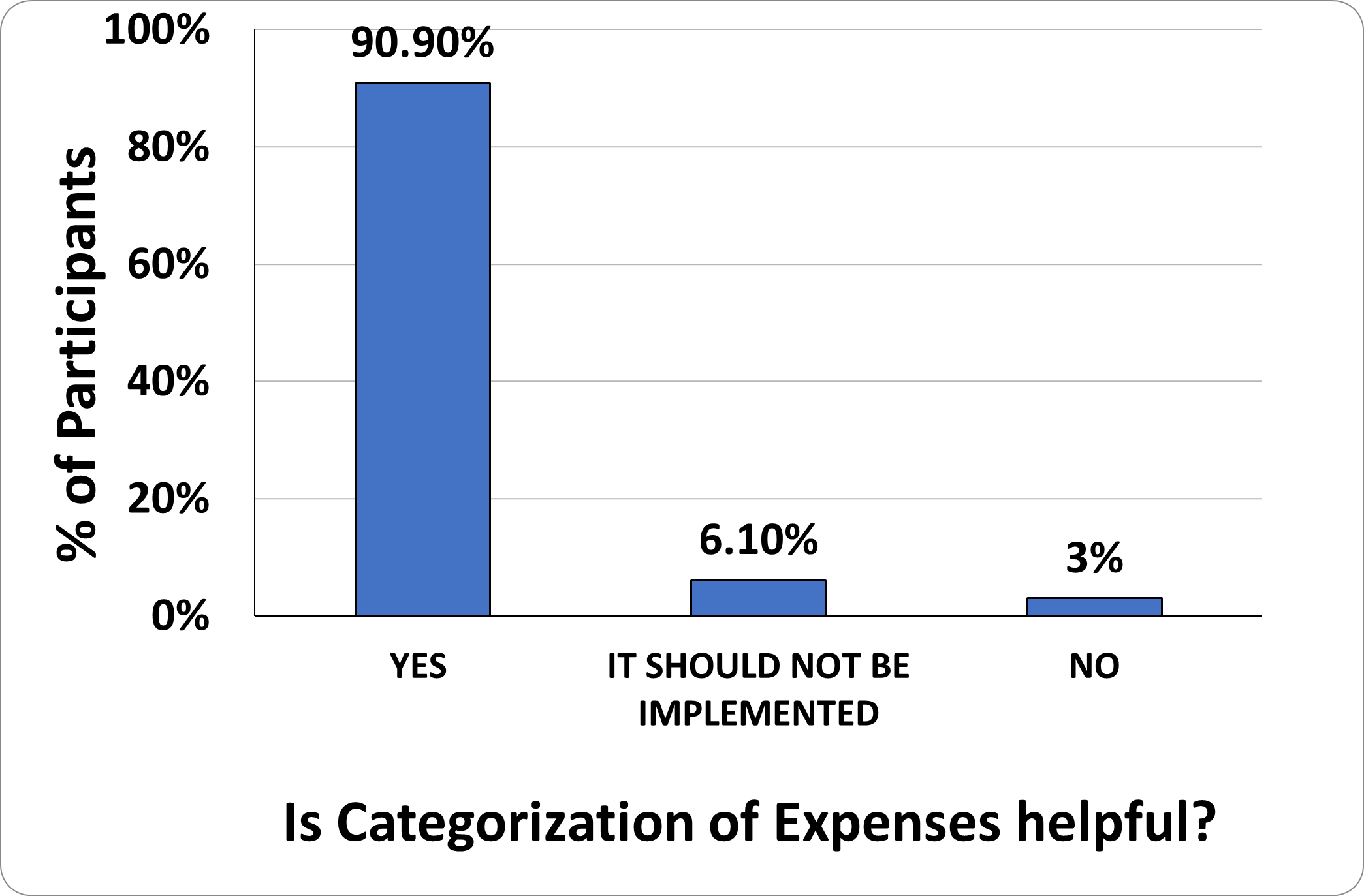}}\hspace{0.05\textwidth}
  \caption{(a) Is incorporating expense Tracker helpful?  (b)  Is categorization of expenses helpful? }
  \label{fig:9a}
  
\end{figure}

The expense tracker feature garnered significant approval as show in Figure \ref{fig:9a}(a)), with 51.5\% of participants rating it as highly beneficial, and 36.4\% finding it beneficial. Meanwhile, 9.1\% remained neutral, and only 3\% considered it not beneficial at all. Overall, 87.9\% of respondents recognized its value, while 12.1\% did not. Furthermore as shown in Figure \ref{fig:9a}(b)), 90.9\% of users found viewing expenses by month and category extremely helpful, with only 6.1\% expressing reservations about its implementation. \\
Regarding payment labeling/categorisation as shown in Table \ref{table:use_cases}, users expressed a clear preference for simplicity and automation. Specifically, 53.1\% of respondents favored simplifying the labeling process to reduce unnecessary steps, while 50\% supported the idea of automatically categorizing payments based on transaction history, suggesting a strong desire for convenience.
A significant portion (43.8\%) also preferred having the option to quickly select from frequently used categories, emphasizing the need for quick and easy selection. The option to skip labeling for fast transactions was favored by 25\% of users, highlighting the importance of flexibility in user experience. Additionally, 40.6\% of respondents felt that using icons or visuals to represent different payment categories would make the process more intuitive, showing a preference for more visually engaging interfaces. Other features, such as drag-and-drop categorization (21.9\%) and voice commands (9.4\%), garnered less support, indicating that users might not prioritize these more advanced functionalities. \\

\begin{figure}[h!]
  \centering
  \subfloat[]{\includegraphics[width=0.4\textwidth]{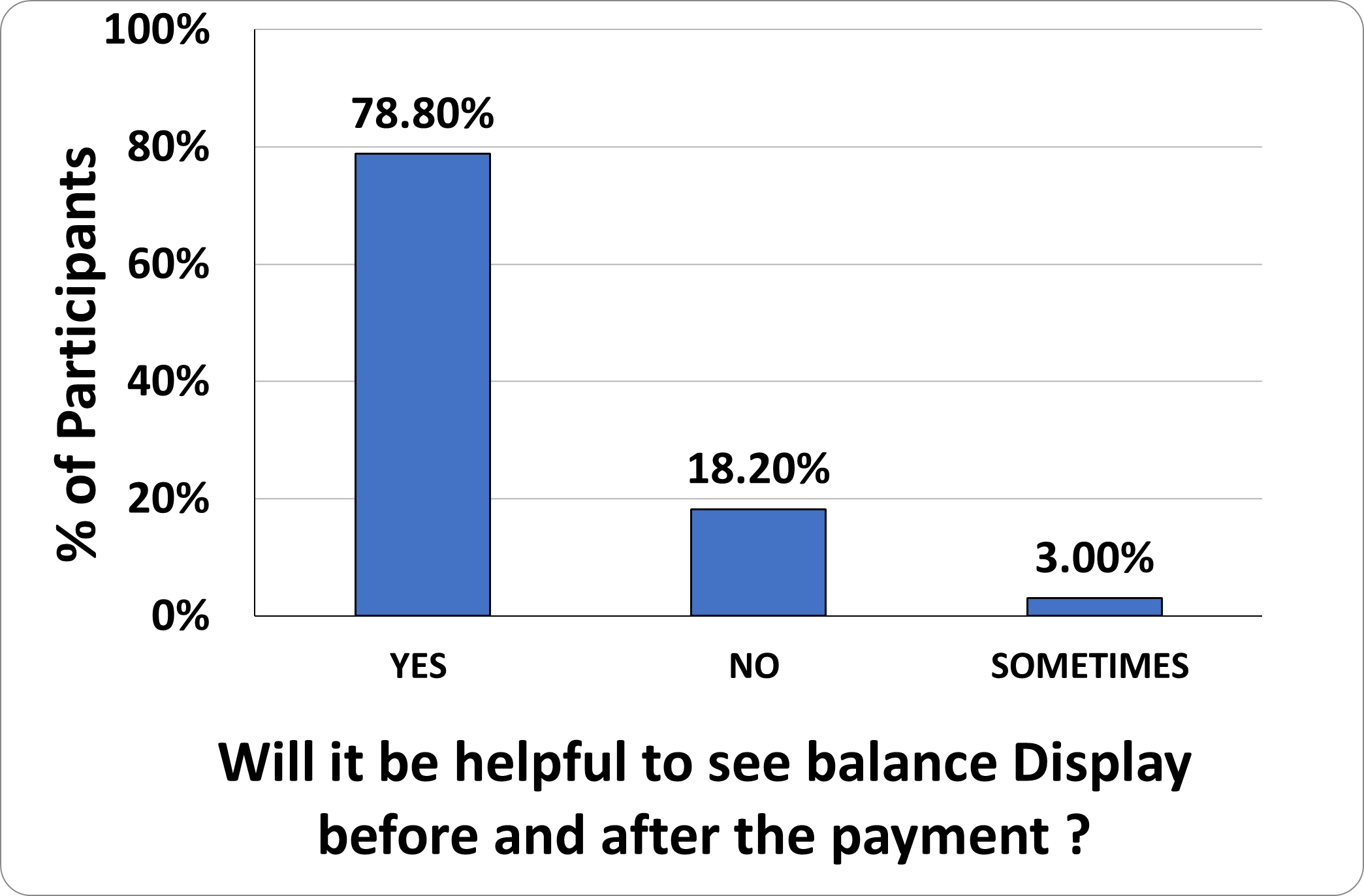}}\hspace{0.05\textwidth}
  \subfloat[]{\includegraphics[width=0.4\textwidth]{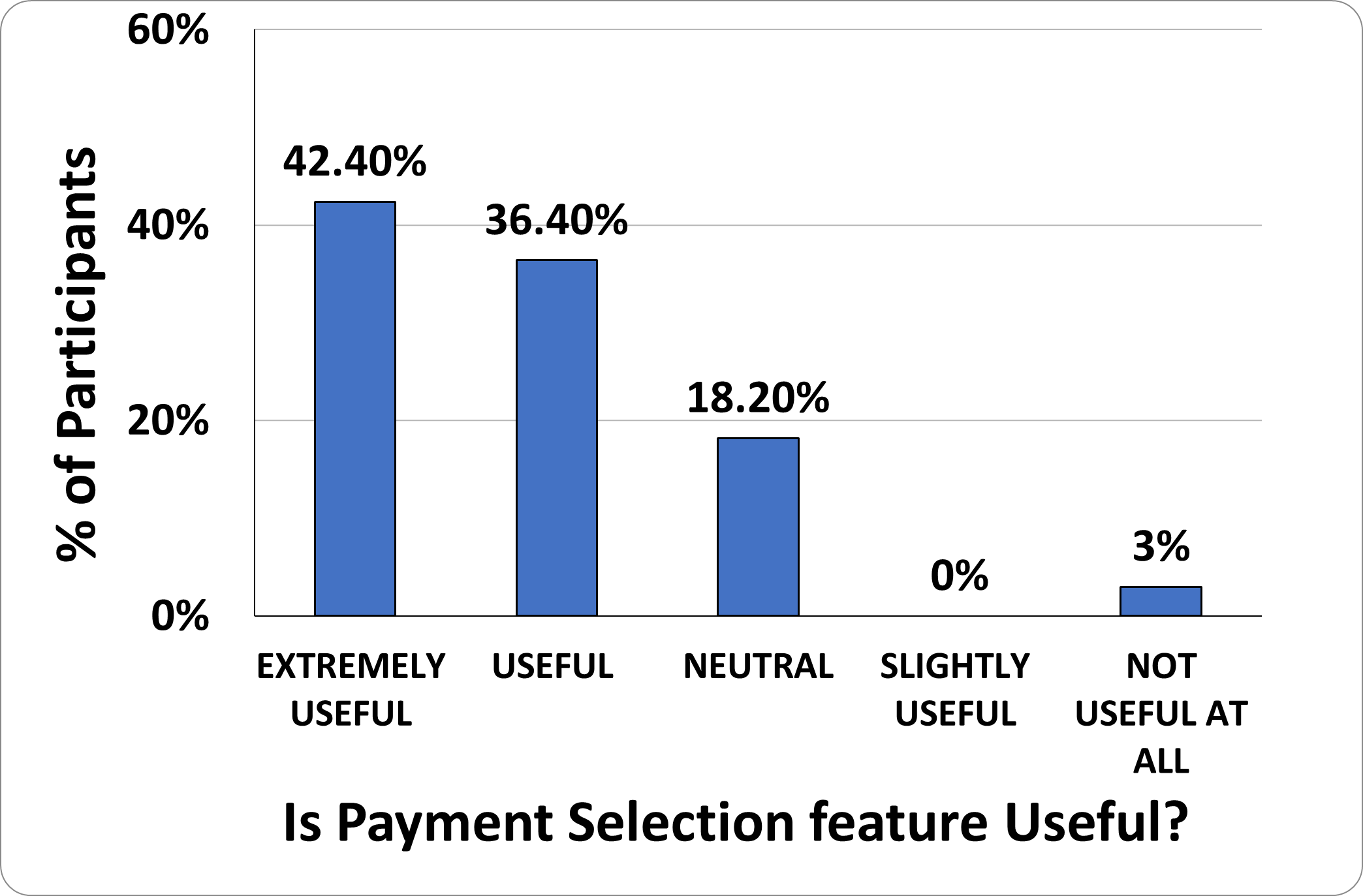}}\hspace{0.05\textwidth}
  \caption{(a) Will it be helpful to see the balance before and after the payment?  (b)  Would daily financial tips be helpful to users? }
  \label{fig:10a}
  
\end{figure}

\begin{figure}[h!]
  \centering
  \includegraphics[width=0.4\textwidth]{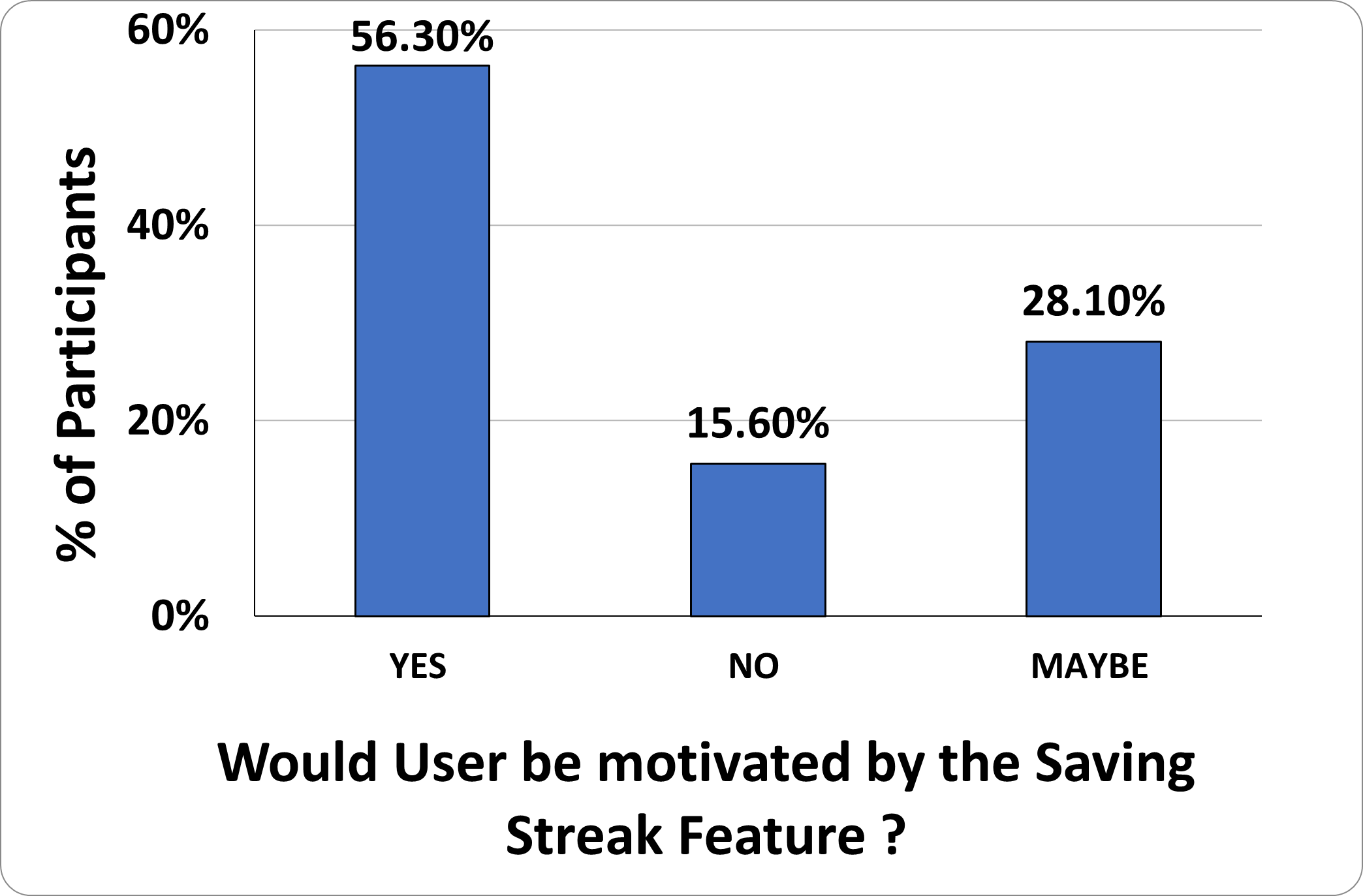}
  \caption{ Would User be motivated by Saving Streak Feature ?}
  \label{fig:11}
\end{figure}

 When it comes to the balance display as shown in Figure \ref{fig:10a}(a)) , 78.8\% of users responded with a definite "yes," indicating they found it consistently helpful to see their balance before and after making a payment. 3\% responded with "sometimes," finding it useful in certain scenarios, while 18.2\% responded with "no," stating they did not find it helpful.  \\
The financial education feature as shown in Table \ref{table:financial_tips} revealed that 21.9\% of users believed the one-liner financial tips might help them spend less and better manage their finances. 56.3\% considered the tips informative but felt they may not significantly impact their spending habits. Meanwhile, 12.5\% thought the tips provided little value and were unlikely to assist with financial management. 6.3\% found the tips distracting and not likely to influence their spending behaviors, while 3.1\% were unable to provide a clear answer. \\
In terms of budgeting limitations as shown in Table \ref{table:budget}, 54.5\% of respondents believed setting budget limits for specific categories would help them manage expenses, while 30.3\% were uncertain about its effectiveness. 12.1\% thought it wouldn't make a significant difference in their spending habits, and 3\% felt it wouldn't help with managing spending or reducing guilt. \\
The payment selection type feature as shown in  Figure \ref{fig:10a}(b)) received favorable feedback, with 78.8\% of users finding it useful. Specifically, 42.4\% felt it was extremely useful for managing expenses, while 36.4\% found it very useful but suggested it could be streamlined. However, 18.2\% rated it as neutral , highlighting areas that may need improvement.\\
Conversely, the saving streak feature as shown in Figure \ref{fig:11}, intended to incentivize saving, was positively received by 56.3\% of users, while 28.1\% remained uncertain, and 15.6\% felt rewards would not influence their saving habits. \\
Users also expressed interest in additional gamification as shown in Table \ref{table:gamification}, with 61.3\% favoring boosted cashback rewards, 45.2\% endorsing random bonuses, and 41.9\% appreciating challenges with rewards for sticking to budgeting goals. When exploring additional gamification features, a majority of 61.3\% supported the introduction of boosted cashback rewards for specific transactions or merchants, reflecting the widespread appeal of tangible financial benefits. Random bonuses or discounts awarded for regular transactions also resonated well, with 45.2\% of respondents endorsing this feature. Challenges that reward users for sticking to budgeting goals were appreciated by 41.9\% of users, signaling a desire for rewards tied to long-term financial discipline. Exclusive, personalized offers based on spending patterns were preferred by 35.5\%, while timely bill payment competitions and consecutive transaction incentives garnered a more moderate appeal, each with 25.8\%. Although interactive scenarios to improve financial knowledge received a lower response rate at 19.4\%, they still provide an opportunity for further development. It’s clear that users are more inclined towards features that directly tie financial rewards or discounts to their everyday spending, while challenges and interactive elements offer additional but secondary appeal. \\
However, the delay feature as shown in Table \ref{table:delay} was less popular. 33.3\% of users found it unnecessary, and another 12.1\% believed it wouldn't impact their spending decisions. Although 24.2\% thought it could help avoid impulsive spending, a larger portion, 30.3\%, was unsure if they would use it frequently. This highlights the need for a more thoughtful integration of such features, considering their varied reception. \\

\subsection{\textbf{Graphs for Quantitative Results}}

\begin{table*}[t]
	\vspace{-1em}
    
	\begin{tabular}{|p{8cm}|p{2cm}| } 
		\hline
		\textbf{Use-Cases} & \textbf{\% of Students}\\
		\hline
		Simplify the labeling process to reduce unnecessary steps& 53.1 \%\\ 
		\hline
        Automatically categorize payments based on transaction history& 50 \%\\ 
		\hline
	  Allow users to quickly select from frequently used categories& 43.8 \%\\
        \hline
        Provide an option to skip labeling for quick transactions& 25 \%\\
        \hline
        Use icons or visuals to represent different payment categories& 40.6 \%\\
        \hline
        Implement a drag-and-drop feature for categorizing expenses& 21.9 \%\\
        \hline
        Enable voice commands to assign and label payments& 9.4 \%\\
        \hline
        Provide personalized suggestions based on previous spending behavior& 6.3 \%\\
        \hline
        Not able to Answer & 3.1 \%\\
        \hline
	\end{tabular}
    \vspace{1em}
	\caption{\textbf{User Preferences For Enhancing The Labeling Process }}
	\label{table:use_cases}
\end{table*}

\begin{table*}[t]
	\vspace{-1em}
    
	\begin{tabular}{|p{8cm}|p{2cm}| } 
		\hline
		\textbf{Use-Cases} & \textbf{\% of Students}\\
		\hline
		Boosted cashback rewards for specific types of transactions or merchants.& 61.3 \%\\ 
		\hline
        Incentives for making consecutive payments or transactions.& 45.2 \%\\ 
		\hline
	  Competitions that reward users for timely bill payments.& 41.9 \%\\
        \hline
        Challenges that encourage sticking to budgeting goals with rewards.& 35.5 \%\\
        \hline
        Exclusive, personalized offers based on spending patterns.& 25.8 \%\\
        \hline
        Interactive scenarios to improve financial knowledge with immediate rewards.& \\
        \hline
        Random bonuses or discounts awarded for regular transactions.& 19.4 \%\\
        \hline
        Not able to answer.& 3.2 \%\\
        \hline
      
	\end{tabular}
    \vspace{1em}
	\caption{\textbf{Preferred Gamification Features for the App }}
	\label{table:gamification}
\end{table*}

\begin{table*}[t]
	\vspace{-1em}
    
	\begin{tabular}{|p{8cm}|p{2cm}| } 
		\hline
		\textbf{Use-Cases} & \textbf{\% of Students}\\
		\hline
		Yes, it helps avoid impulsive spending.& 24.2 \%\\ 
		\hline
        Yes, might be useful, but unsure about use.& 30.3 \%\\ 
		\hline
	  No, seems unnecessary and inconvenient.& 33.3 \%\\
        \hline
        No, won’t impact spending decisions.& 12.1 \%\\
        \hline
        Not able to answer.& 0 \%\\
        \hline
	\end{tabular}
    \vspace{1em}
	\caption{\textbf{
Would It Be Helpful for Users to Have a Timer (e.g., 8-10 Seconds) Before Completing a Large Payment, Allowing Time to Reconsider?     }}
	\label{table:delay}
\end{table*}

\begin{table*}[t]
	\vspace{-1em}
    
	\begin{tabular}{|p{8cm}|p{2cm}| } 
		\hline
		\textbf{Use-Cases} & \textbf{\% of Students}\\
		\hline
		Yes, it will help me spend less and manage finances.& 54.5 \%\\ 
		\hline
        Yes, it might help but I’m not sure how effective.& 30.3  \%\\ 
		\hline
	  No, it likely won’t significantly affect spending.& 12.1 \%\\
        \hline
        No, it won’t help with spending or guilt.& 3 \%\\
        \hline
        Not able to answer.& 0
\\
        \hline
	\end{tabular}
    \vspace{1em}
	\caption{\textbf{
Would Setting Budget Limits for Categories Like Food or Travel Help Manage Spending?    }}
	\label{table:budget}

\end{table*}

\begin{table*}[t]
	\vspace{-1em}
    
	\begin{tabular}{|p{8cm}|p{2cm}| } 
		\hline
		\textbf{Use-Cases} & \textbf{\% of Students}\\
		\hline
		Yes, they help me spend less and manage finances.& 21.90%
\\ 
		\hline
        Yes, they're informative but don’t significantly impact spending.& 56.30%
\\ 
		\hline
	  No, they don't provide much value for financial management.& 12.50%
\\
        \hline
        No, they're distracting and unlikely to influence spending.& 6.30%
\\
        \hline
        Not able to answer.& 3.10%
\\
        \hline
	\end{tabular}
    \vspace{1em}
	\caption{\textbf{
Would Short Quotes on Financial Management, Budgeting Tips, and Related Facts Be Helpful ?}}
	\label{table:financial_tips}
\end{table*}

\begin{table*}[t]
	\vspace{-1em}
    
	\begin{tabular}{|p{8cm}|p{2cm}| } 
		\hline
		\textbf{Use-Cases} & \textbf{\% of Students}\\
		\hline
		Extremely useful& 42.40%
\\ 
		\hline
        Useful& 36.40%
\\ 
		\hline
	  Neutral& 18.20%
\\
        \hline
        Slightly useful& 0%
\\
        \hline
        Not useful at all& 3%
\\
        \hline
	\end{tabular}
    \vspace{1em}
	\caption{\textbf{
Is It Useful to Label the Payment Type (e.g., Food, Travel) Before or After a Transaction?     }}
	\label{table:label}
\end{table*}

\subsection{Qualitative Analysis}
The thematic analysis of 20 participant interviews revealed valuable insights into the strengths and areas for improvement of the UPI app prototype. The participants provided feedback on various features such as navigation, data representation, impulse control, and saving motivation, leading to the identification of ten key themes. These themes capture the overall user experience and highlight recommendations for future refinements. Table \ref{table:thematic_analysis} presents the summary of the identified themes while the details are as follows:

\subsubsection{\textbf{Motivation to Save}}\hfill\\
The saving streak feature, which rewards users for saving, was positively received by many participants. P10 commented, \emph{\textbf{"The saving streaks would definitely motivate me to save more, especially with rewards as an incentive."}} Additionally, P9 suggested enhancing this feature by adding a leaderboard: \emph{\textbf{"A leaderboard for saving streaks would be cool to compete with friends."}}
While many participants found gamification motivating, others, like P14, felt it should be optional: \emph{\textbf{"The gamification elements are fun, but not everyone might be into them, so having an option to turn them off could be cool."}}
Moreover, P15 highlighted that setting savings goals was particularly motivating:\emph{\textbf{ "Setting savings goals is super motivating—I’m excited to see how much I can save each month."}}

\subsubsection{\textbf{Addressing Overspending and UPI Payments}}\hfill\\
Several participants appreciated the app’s focus on helping users manage their spending through UPI payments. P2 highlighted the app’s thoughtful approach: \emph{\textbf{"I appreciated the overall idea of addressing overspending with UPI payments."}}
The payment type selection feature before making payments was also seen as beneficial. P9 found it especially helpful, stating, \emph{\textbf{"The payment type selection feature before making a payment is super helpful."}}

\subsubsection{\textbf{Data Representation in Graphs}}\hfill\\
The app’s use of visual representations, such as bar graphs for tracking expenses, was generally well-received. P3 appreciated the feature, stating, "The bar graphs for analyzing expenditures over the past three months were particularly useful." However, P1 suggested a refinement: \emph{\textbf{ "Displaying percentages on the bar graphs instead of absolute amounts would make data interpretation easier."}}
Additionally, some participants desired more comparative data features. P8 proposed, \emph{\textbf{"A graph showing category-wise comparisons over several months would be beneficial to track anomalies and shifts in spending patterns." }}
By enhancing the flexibility and clarity of data representation, the app can provide users with better insights into their spending habits.

\subsubsection{\textbf{Prototype Usability and User Experience}}\hfill\\
Participants found the prototype helpful overall but suggested refinements to enhance its usability. P2 expressed a positive experience but noted, \emph{\textbf{"Improving aesthetics and refining rewards for a better user experience would be helpful." }} Similarly, P3 highlighted how the app’s design was thoughtful and practical for managing finances.

However, some features were seen as restrictive. P5 remarked, \emph{\textbf{"Some features might be cumbersome, especially if they delay transactions when enabled." }} Others, like P6, found certain elements unnecessary, such as the saving streak feature: \emph{\textbf{"It might be unnecessary for me but motivating for others."}}

Participants emphasized that the app might be more relevant to students or new users, as P5 pointed out: \emph{\textbf{"The prototype might be more relevant for individuals new to managing expenses through UPI." }} Flexibility to enable or disable certain features based on user preference would allow the app to cater to a broader audience.

\subsubsection{\textbf{Payment Efficiency}}\hfill\\
Participants also provided feedback on improving the efficiency of payments. P17 and P7 both found the process of typing a PIN multiple times redundant, with P17 commenting, \emph{\textbf{"Typing PINs twice is not necessary; it just causes delays." }}

In addition, P3 suggested adding a feature to save frequently used QR scanners to streamline payments: \emph{\textbf{"It would be great to save frequently used QR scanners to make the payment process faster."}}
These insights point to the need for a more seamless payment process, especially for frequent users.

\subsubsection{\textbf{Financial Education and Thoughtful Design}}\hfill\\
The financial education pop-ups received mixed feedback. While some, like P19, found them informative, she noted, \emph{\textbf{ "They may not significantly impact my spending habits." Similarly, P9 commented, "They might help, but I’m not sure how effective they will be."}}
To reduce intrusiveness, participants suggested giving users more control over these pop-ups. For instance, adding an option to reduce the frequency of pop-ups or turn them off during transactions could improve the experience for users who want quick and seamless payments.

\subsubsection{\textbf{Aesthetic Appeal and Visual Design}}\hfill\\
The app’s visual design was praised for its color scheme, but several participants suggested ways to enhance the aesthetic experience. P18 mentioned, \emph{\textbf{ "The UI can be more minimalistic and seamless," while P17 felt, "Design layout could be a little better, but overall everything is great." }}
Some participants, such as P7, suggested using colors associated with finance, like green, to improve the app’s visual appeal. Enhancing the aesthetic design by making it more minimalistic and relevant to the financial context could improve the user experience.

\subsubsection{\textbf{Impulse Control (Delay Feature)}}\hfill\\
The delay feature, designed to help users manage impulse spending, received mixed reactions. P1 supported the feature, saying, \emph{\textbf{"It would help me make more deliberate decisions and avoid impulsive spending." However, others found it inconvenient.}} P3 mentioned, \emph{\textbf{"No, it seems unnecessary and could be an inconvenience."}}

For some, such as P19, the delay feature was valuable: \emph{\textbf{"The large-amount feature timer is good for me." }} However, others like P5 found it potentially problematic if it delayed transactions, highlighting that it could be more selectively applied based on transaction type.

\subsubsection{\textbf{Navigation Clarity}}\hfill\\
Several participants experienced difficulties with the navigation system, particularly when switching between different views in the expense tracker. P1 mentioned, "Users might not easily understand how to switch between views or view daily expenses,\emph{\textbf{" emphasizing that clearer instructions or features are needed to enhance usability.}} P19 suggested a practical improvement by proposing, \emph{\textbf{"It would be great if we have another button for switching payment month-wise, so users can locate it easily."}}
This feedback suggests that adding more intuitive and discoverable navigation features would significantly improve the user experience, especially for switching between calendar and month views.\\
\subsubsection{\textbf{User-Centric Design and Behavior Focus}}\hfill\\
Participants emphasized the importance of designing the app based on observed user behavior, rather than assumptions. P4 stressed, \emph{\textbf{"It’s crucial to validate whether users actually need or use such features and how frequently they check their expenses."}}
This feedback highlights the need for thorough user testing to ensure the app’s features align with actual user habits and needs. By conducting behavior-based research, the app can better support user adoption and engagement.

\begin{table*}[t]
	\vspace{-1em}
    
	\begin{tabular}{|p{4.5cm}|p{4.5cm}|p{4.5cm}|} 
		\hline
		 \textbf{Theme}&\textbf{Description}& \textbf{Quotation (with Participant ID)}\\
 & &\\\hline
		 \textbf{Motivation to Save}&Focuses on how features like saving streaks and rewards motivate users to save more.& "The saving streaks would definitely motivate me to save more, especially with rewards as an incentive." (P10)
\\ 
		\hline
         \textbf{Addressing Overspending and UPI Payments}&Explores how the app helps users manage their spending with UPI payments.& "I appreciated the overall idea of addressing overspending with UPI payments." (P2)
\\ 
		\hline
	   \textbf{Data Representation in Graphs}&Discusses user feedback on the clarity and effectiveness of the app’s graphs and charts.& "Displaying percentages on the bar graphs instead of absolute amounts would make data interpretation easier." (P1)
\\
        \hline
         \textbf{Prototype Usability and User Experience}&Covers overall user experience, highlighting useful features and areas for improvement.& "The prototype might be more relevant for individuals new to managing expenses through UPI." (P5)
\\
        \hline
         \textbf{Payment Efficiency}&Focuses on suggestions to streamline payment processes, including simplifying PIN entry.& "Typing PINs twice is not necessary; it just causes delays." (P17)
\\
        \hline
         \textbf{Financial Education and Thoughtful Design}&Reflects on how financial education pop-ups can help users manage spending habits.& "They may not significantly impact my spending habits." (P19)
\\
        \hline
         \textbf{Aesthetic Appeal and Visual Design}&Examines user feedback on the app’s visual design and color schemes.& "The UI can be more minimalistic and seamless." (P18)
\\
        \hline
         \textbf{Impulse Control}&Analyzes how features like the delay timer help or hinder impulse control.& "It would help me make more deliberate decisions and avoid impulsive spending." (P1)
\\
        \hline
         \textbf{Navigation Clarity}&Focuses on user challenges with navigating between different views in the app.& "Users might not easily understand how to switch between views or view daily expenses." (P1)
\\
        \hline
 \textbf{User-Centric Design and Behavior Focus}& Highlights the importance of designing the app based on real user behavior and needs.&"It’s crucial to validate whether users actually need or use such features and how frequently they check their expenses." (P4)\\
 \hline
	\end{tabular}
    \vspace{1em}
	\caption{\textbf{Thematic Analysis}}
	\label{table:thematic_analysis}
\end{table*}

\section{Phase 2: Incorporating User Recommendations from Usability Testing }\label{sec:incorportating-user-recommendations}

\begin{figure*}
    \centering
    \includegraphics[width=1\linewidth]{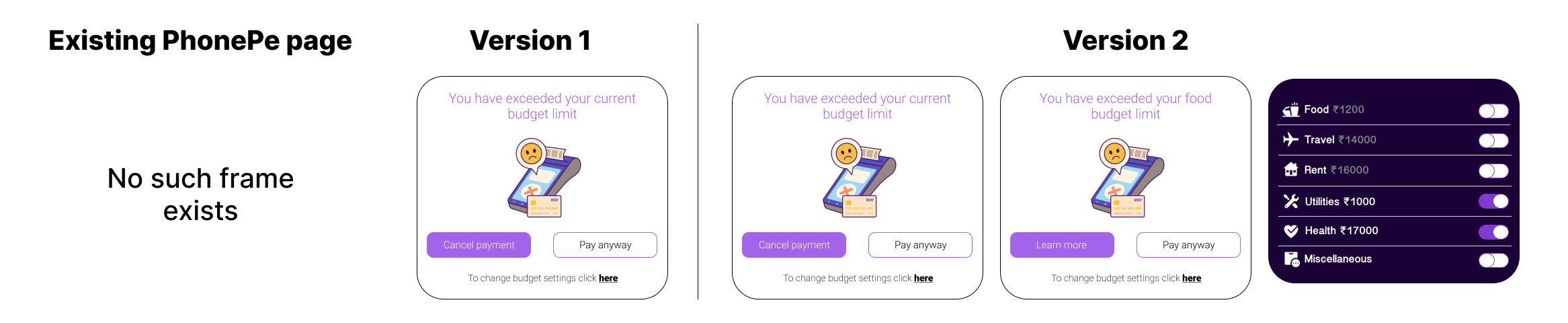}
    \caption{\textbf{Budget Limitations and Pop Ups}}
    \label{fig:limitation}
\end{figure*}

Several key features were introduced to enhance financial management and organisation, as illustrated by the existing PhonePe page, Version 1, and Version 2 for each feature . The budget and limitations, as shown in Figure \ref{fig:limitation}, a pop-up feature was created due to the absence of any mechanism for users to set or track spending limits in the original app. In version 1, a pop-up was implemented to notify users when their budget was exceeded, offering three options: proceed with the payment, adjust the budget in the settings, or cancel the transaction. After feedback from users, version 2 incorporated a new feature allowing budget limits to be set for individual expense categories, providing more granular control over finances.
The categorization of expenses was another feature developed, as shown in Figure \ref{fig:expense category}, as the original app did not allow users to organize expenses into categories. Version 1 introduced a page where expenses could be categorized under predefined headings such as Travel, Rent, and Food. Based on user feedback from usability studies, a "Miscellaneous Expenses" category was added in version 2 to offer greater flexibility for categorizing unclassified expenses.
Regarding the transaction history, as shown in Figure \ref{fig:transaction history}, the existing app displayed transactions but lacked color-coding and the ability to edit categories. Version 1 introduced color codes—red for credited transactions and green for debited ones—to make transaction history more visually intuitive. Based on user feedback from usability studies, version 2 added an "edit category" option, allowing users to modify the category of a transaction even after it has been recorded.
The settings page was redesigned to offer greater control over financial management features as shown in Figure \ref{fig:settings}. Initially, it only provided basic options like theme and data preferences. In version 1, the settings were made more customizable, allowing users to opt in or out of features such as budget limitations, expenditure tracking, and large payment notifications. 

\begin{figure*}
    \centering
    \includegraphics[width=1\linewidth]{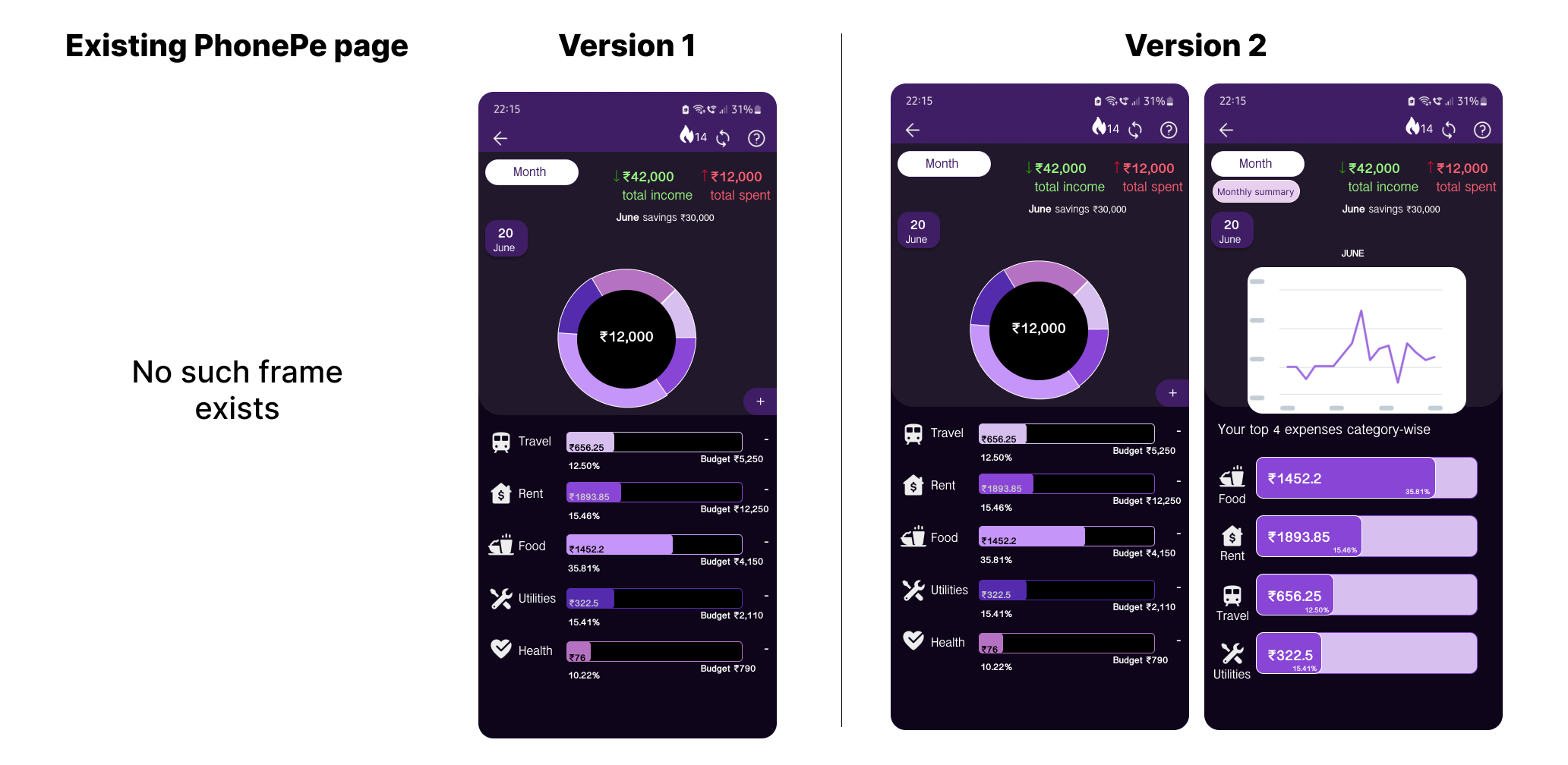}
    \caption{\textbf{Dashboard}}
    \label{fig:dashboard}
\end{figure*}

\begin{figure*}
    \centering
    \includegraphics[width=1\linewidth]{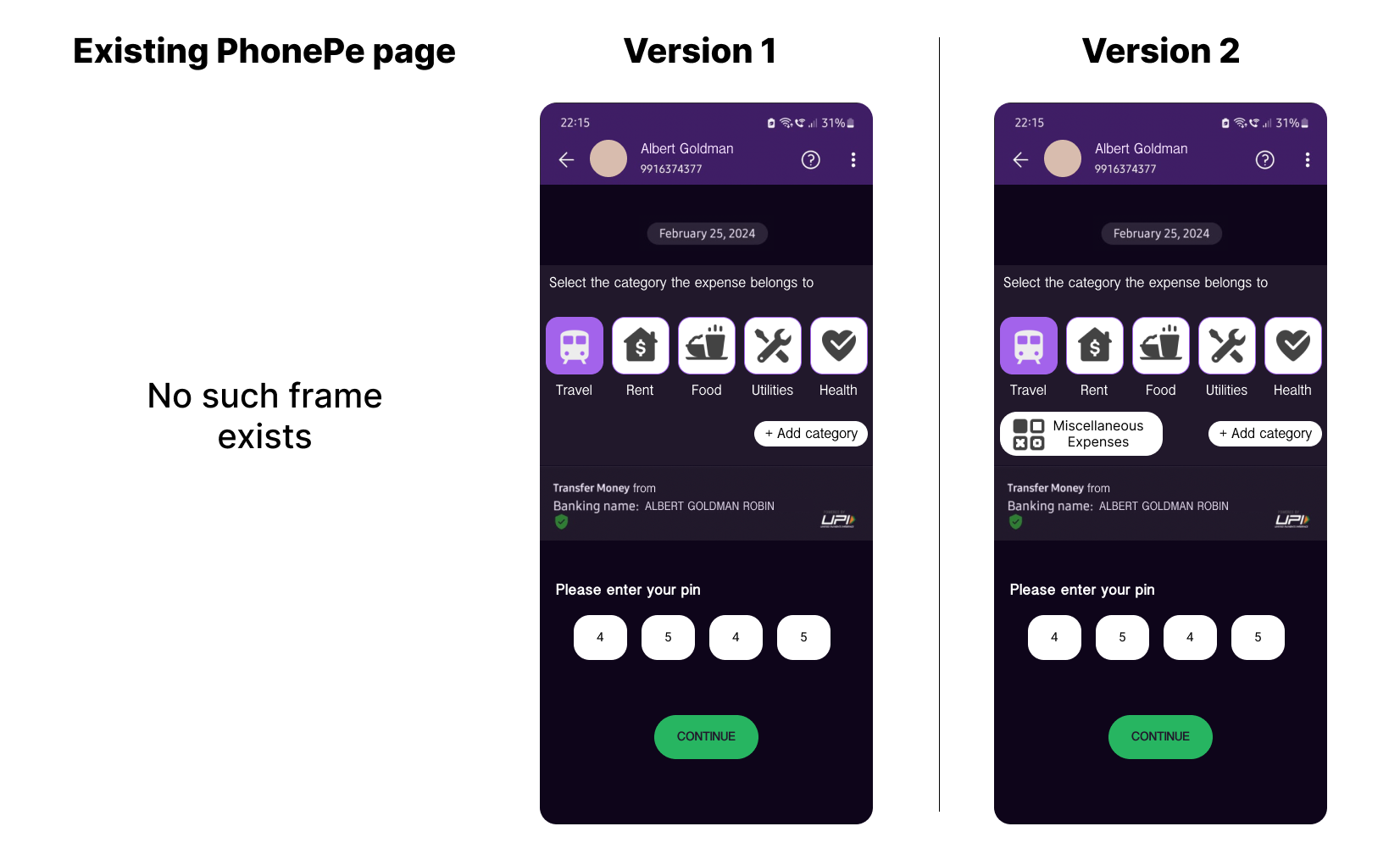}
    \caption{\textbf{Categorization of Expenses}}
    \label{fig:expense category}
\end{figure*}

\begin{figure*}
    \centering
    \includegraphics[width=1\linewidth]{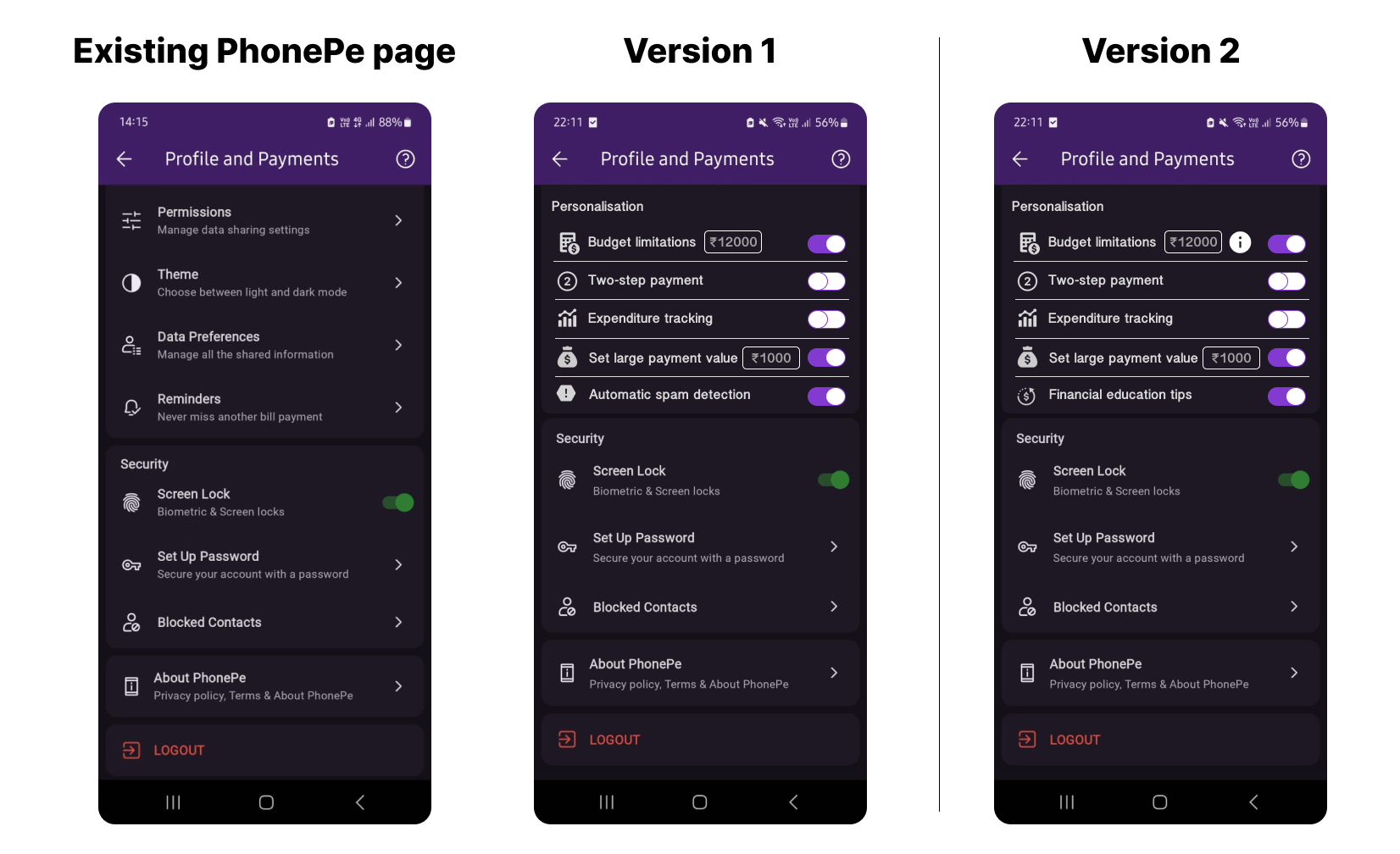}
    \caption{\textbf{Settings}}
    \label{fig:settings}
\end{figure*}

\begin{figure*}
    \centering
    \includegraphics[width=1\linewidth]{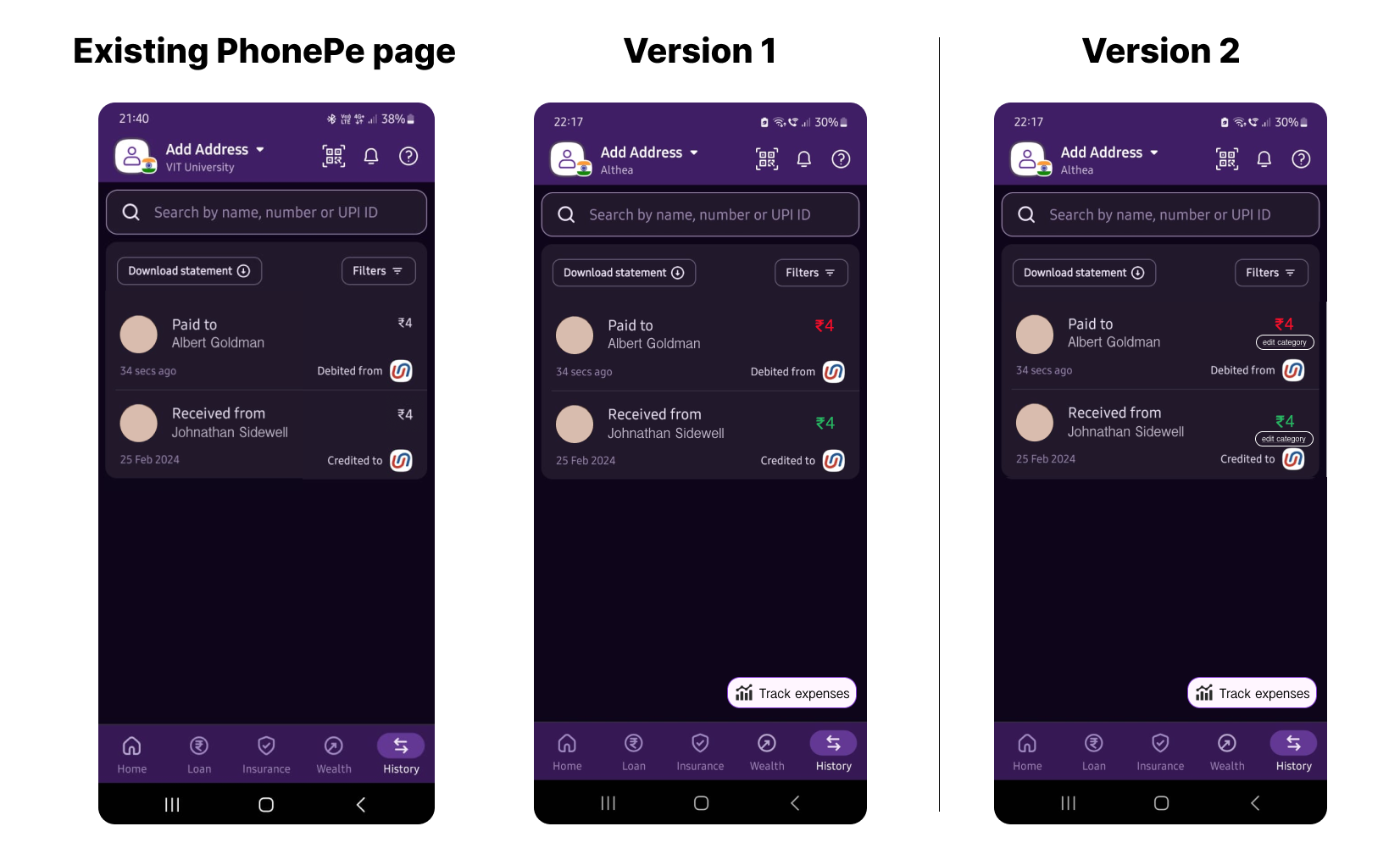}
    \caption{\textbf{Transaction History}}
    \label{fig:transaction history}
\end{figure*}

Based on usability studies, the "automatic spam detection" feature was replaced with "financial education tips" in version 2, as it was deemed more relevant to users' needs.
The dashboard feature as shown in Figure \ref{fig:dashboard}, which did not exist in the original app, was introduced in version 1 to give users an overview of their financial activity. This included daily and monthly summaries of expenditures by category, with percentage breakdowns, total expenses, and savings. After further testing, version 2 was enhanced with a spending summary that featured a graph of monthly expenditure trends and highlighted the top four categories where the most money was spent, offering users more valuable insights into their financial behaviour.
Based on usability testing and user feedback, several critical improvements were implemented in the final prototype to enhance its functionality and user experience. One of the main recommendations was the inclusion of a 'Miscellaneous' category for expenses, which was previously absent. This category allows users to track undefined or irregular expenses more efficiently, addressing gaps identified in the initial prototype. \\
Another significant change involved adding the ability to modify the category of a transaction directly from the history page. This feature was introduced to ensure greater flexibility and accuracy in classifying expenditures as per user’s personal understanding. It was found that users often miscategorized expenses and needed a way to make corrections.\\
The home screen underwent a substantial update, with the addition of a display showing the top five spending categories. This feature provides users with immediate insights into their most frequent spending areas, offering a more focused and intuitive overview of their financial behavior. At the same time, the monthly view was enhanced to show a complete breakdown of spending across all categories, offering a comprehensive analysis of monthly expenditures that was not available in the earlier version.\\
To facilitate better tracking of spending trends, a line graph feature was introduced, enabling users to compare category-wise expenditures across months. This graphical representation helps users visually assess their spending habits and correct themselves accordingly if needed. \\
Additionally, the ability to export statements was included, providing users with an easy way to generate and share their financial data, which was not a part of the initial design. \\
Furthermore, category-wise spending limitations were introduced in the final version. This feature alerts users when their spending in a particular category, such as food, exceeds a predefined limit, encouraging better budgeting and financial discipline. \\
Although some features, such as the saving streak functionality, remained as recommendations rather than full implementations, the final prototype integrated these critical user-driven improvements, resulting in a more polished and practical financial management tool. These changes significantly enhance the app's utility in real-world financial management scenarios.

\section{Recommendations for UPI App Developers and Various Stakeholders}\label{sec:recommendations-for-developers}
Our extensive user testing and feedback gathered valuable insights and feature recommendations for improving UPI apps. We implemented some user recommendations and verified them through usability testing. However, it was only feasible to implement some suggestions. The following key recommendations guide UPI app developers, product managers, and involved stakeholders to make UPI applications more user-friendly, intuitive, and engaging. These recommendations focus on enhancing user experience, personalizing interactions, and optimizing financial management within the app. The technical implementation details are left to the discretion of UPI apps' product managers and developers, considering their specific constraints and business requirements.

The critical insight from our study is that users overwhelmingly prefer a single, integrated app to manage most of their financial needs rather than switching between multiple platforms. Given that UPI apps handle users' most sensitive data—their financial transactions—there is a prime opportunity to implement the following recommended features. 

\subsection{Recommended Features}

\subsubsection{\textbf{Expenses Tracker}}
We recommend integrating an inbuilt expense tracker to help users monitor their spending patterns. This feature should allow users to categorize their expenses (e.g., food, travel, entertainment, etc.) and provide daily and month summaries of incoming and outgoing cash flows. By offering clear insights into their spending habits, users will be empowered to set personal budgets, compare trends, and make more informed financial decisions.

\subsubsection{\textbf{Payment Category Selection}}
Another recommendation is to incorporate a payment category selection option, which allows users to categorize their expenses before making payments. This feature will feed directly into the expense tracker, giving users a clear breakdown of where their money is being spent and highlighting areas where they may be overspending. .

\subsubsection{\textbf{Automatic and Flexible Transaction Category Tagging}}
Automatically tagging transactions based on parameters such as the merchant’s business type or user history would alleviate the need for manual tagging. Additionally, a system allowing transactions to belong to multiple categories simultaneously (e.g., a single transaction covering entertainment and food) would better reflect real-world spending. Users should still be able to adjust tags if necessary when transactions are automatically miscategorized manually. An ample selection of categories should also cover diverse spending areas.

\subsubsection{\textbf{Category-Wise Budget Exceptions}}
While users liked category-wise spending limits, there was concern about making payments that exceeded a limit. A solution could be to allow "exceptional" payments that exceed the category limit, with an option to flag these exceptions for future reference or analysis.

\subsubsection{\textbf{Data Visualization for Spending Insights}}
Representing users’ spending habits using intuitive visualizations such as bar graphs or pie charts can help them draw insights and make better financial decisions. Minimalism and personalization should be critical elements in these representations. Improving the calendar view to show daily and monthly expenses would enhance users' understanding of their financial patterns. Additionally, the app could track recurring payments and asset portfolios to provide a comprehensive view of personal finances, thus minimizing the need for users to rely on multiple apps.

\subsubsection{\textbf{Saving Streaks and Rewards} }
A “saving streak and rewards” feature was highly appreciated during testing, as it incentivizes users to stick to their financial goals. Potential implementations include challenges tied to budgeting and offering rewards such as cashback, discounts, or bonuses for consistent saving. A leaderboard comparing users' saving streaks with their peers could provide motivation and enhance engagement.

\subsubsection{\textbf{Daily Expense Summaries} }
Giving users a daily summary of their expenses, including notable trends or anomalies, would increase their financial awareness and help them monitor their budgets more closely.

\subsubsection{\textbf{AI-Powered Financial Insights}}
With the growing use of AI, incorporating AI-driven financial insights into UPI apps could help users manage their expenses more efficiently. AI could analyze spending trends, offer personalized tips, and predict future financial behaviors. However, developers must address user privacy concerns to protect sensitive financial data.

\subsubsection{\textbf{Displaying Balance Before and After Payments}}
To make users more conscious of their spending, displaying their account balance both before and after a transaction would be useful. This reinforces financial awareness and can help curb unnecessary spending.

\subsubsection{\textbf{Financial Education and Awareness}}
To promote responsible financial behaviour, UPI apps could collaborate with financial education platforms and stakeholders. By integrating financial literacy resources, users would have easy access to tools that help them be better educated with respect to finance and make informed decisions, fostering better financial habits over time.

\subsubsection{\textbf{Personalized Financial Popups}}
Some users appreciated the concept of financial popups, but to enhance relevance, we recommend personalizing these popups based on user spending habits or profession. For instance, users interested in stock markets could receive popups with market insights. Additionally, incorporating a "skip" or "more information" button will give users control over how much detail they wish to engage with, depending on their immediate context and preferences.

\subsubsection{\textbf{Scam Detection Feature}}
Given the rise in scams, a scam detection feature is essential. It would alert users when making payments to UPI IDs flagged by multiple users as potential scams. This proactive feature safeguards users from fraudulent activities and encourages caution before completing any transactions.

\subsubsection{\textbf{Simulating Payment Friction with Delays}}
Mimicking real-world payment scenarios by adding a slight delay before transaction completion could replicate the natural “thought process” of spending. However, as this was not universally favored by users, we recommend an optional toggle to enable/disable or control the frequency of delayed payments. This customization would allow users to opt in based on personal preferences.

\section{Discussion}\label{sec:discuss}
Answering \textbf{RQ\ref{RQ1}} as mentioned in Section \ref{sec:intro}, the study results reveal a significant impact of UPI on Indian users' spending habits. Most respondents (74.2\%) accepted that UPI adoption led to increased spending, highlighting the UPI influence on user behavior. The data also indicates that many respondents started using UPI during significant economic and social changes, such as demonetization and pandemic-related lockdowns. Furthermore, 59.8\% of respondents acknowledged exceeding their budget or overspending due to UPI usage. The qualitative analysis supports these findings, indicating that that UPI has contributed to increased spending for most users. The abstract nature of digital transactions leads to psychological shifts, contributing to impulsive spending. In some cases, regret and guilt hit people after they realize their spending by checking their bank balances or statements, compared to spending physical cash, highlighting the lack of tangibility associated with UPI payments.

Answering \textbf{RQ\ref{RQ2}} as mentioned in Section \ref{sec:intro}, user satisfaction levels were high, with 91.5\% of the survey respondents reporting satisfaction with UPI usage and 95.2\% of the survey respondents found making payments via UPI convenient. The qualitative analysis underscores the convenience of UPI transactions, with users praising its speed, efficiency, and the elimination of the need for physical cash. The absence of processing fees, as opposed to credit and debit cards, is also highlighted as a positive factor contributing to user satisfaction. The research findings indicate that users perceive UPI as a secure and efficient payment method, eliminating concerns about carrying large sums of money. This positive perception of convenience and efficiency likely contributes to the widespread adoption and satisfaction with UPI, ultimately shaping users' preferences for cashless transactions and influencing spending behavior. 




Answering \textbf{RQ\ref{RQ3}} as mentioned in Section \ref{sec:intro}, the study identifies variations in UPI adoption across demographic groups, particularly in professions. Working professionals and business owners integrate UPI into financial routines seamlessly, while students use it for academic needs. Despite variations, common spending trends are observed across age groups and professions, showcasing UPI's adaptability to diverse user requirements. 


Answering \textbf{RQ\ref{RQ4}}, as per users, current UPI apps, work just fine for enabling them to conduct transactions. Making UPI payments has caused overspending due to the lack of tangibility associated with digital payment and the lack of "friction"/"natural hesitation" while conducting digital transactions compared to physical currency. One common theme that came out of our study is that users prefer having financial tools in one app rather than switching between multiple apps/platforms. Even after making payments themselves, users prefer being aware of their spending habits and patterns. Having tools embedded inside their preferred UPI apps saves the time and cognitive load of juggling between platforms to make sense of their data - financial transactions. Some features of our prototypes were lauded by the users like the expense tracker, category selection etc. The same has been mentioned in Section 11 for UPI app stakeholders to look into and implement at their end. Doing so would lead to high user satisfaction and a better understanding of the user's financial data.

The Version 1 (V1) of the prototype can be found at \href{https://www.figma.com/proto/8rTQ9KxtXpErWL8aSJBk5q/current-projects?node-id=1083-455&node-type=canvas&t=0VUqMHceFZpfs7GY-1&scaling=scale-down&content-scaling=fixed&page-id=0%3A1&starting-point-node-id=1083%3A455}{\color{blue}{this link}}.
The Version 2 (V2) of the prototype can be viewed at \href{https://www.figma.com/proto/fMIdUhq2iHmF1qG6k22odI/New-page?t=TUOQ2Dc8YPqdYF4h-1&scaling=scale-down&content-scaling=fixed&page-id=0%3A1&node-id=393-3&starting-point-node-id=393%3A3}{\color{blue}{this link}}.

\section{Conclusion}\label{sec:conclusion}
This study provides valuable insights into how UPI influences individual spending behavior and highlights user-driven improvements for UPI applications. Our mixed-methods approach revealed that a majority of users report increased spending due to UPI’s intangible nature, while also expressing high satisfaction with its convenience.

We developed a high-fidelity prototype based on user feedback, incorporating features like an expense tracker and budgeting tools. Usability testing confirmed that these features, particularly those promoting financial awareness, were highly valued by users. Our research offers actionable recommendations for UPI app developers and other stakeholders to enhance user engagement and support responsible financial management.



\newpage
\bibliographystyle{ACM-Reference-Format}
\bibliography{chatgpt-ref}


\begin{thebibliography}{71}


\ifx \showCODEN    \undefined \def \showCODEN     #1{\unskip}     \fi
\ifx \showDOI      \undefined \def \showDOI       #1{#1}\fi
\ifx \showISBNx    \undefined \def \showISBNx     #1{\unskip}     \fi
\ifx \showISBNxiii \undefined \def \showISBNxiii  #1{\unskip}     \fi
\ifx \showISSN     \undefined \def \showISSN      #1{\unskip}     \fi
\ifx \showLCCN     \undefined \def \showLCCN      #1{\unskip}     \fi
\ifx \shownote     \undefined \def \shownote      #1{#1}          \fi
\ifx \showarticletitle \undefined \def \showarticletitle #1{#1}   \fi
\ifx \showURL      \undefined \def \showURL       {\relax}        \fi
\providecommand\bibfield[2]{#2}
\providecommand\bibinfo[2]{#2}
\providecommand\natexlab[1]{#1}
\providecommand\showeprint[2][]{arXiv:#2}

\bibitem[Abr(2004)]%
        {Abrazhevich}
 \bibinfo{year}{2004}\natexlab{}.
\newblock \bibinfo{title}{Electronic Payment Systems: A User-Centered Perspective and Interaction Design}.
\newblock
\newblock
\urldef\tempurl%
\url{https://pure.tue.nl/ws/portalfiles/portal/2396269/200411085.pdf}
\showURL{%
\tempurl}


\bibitem[bax(2005)]%
        {baxter}
 \bibinfo{year}{2005}\natexlab{}.
\newblock \bibinfo{title}{Bank interchange of transactional paper; Legal and Economic Perspectives}.
\newblock
\newblock
\urldef\tempurl%
\url{https://neconomides.stern.nyu.edu/networks/phdcourse/Baxter_Bank_interchange_of_transactional_paper.pdf}
\showURL{%
\tempurl}


\bibitem[Lus(2009)]%
        {Lusardo_Tufano}
 \bibinfo{year}{2009}\natexlab{}.
\newblock \bibinfo{title}{DEBT LITERACY, FINANCIAL EXPERIENCES, AND OVERINDEBTEDNESS}.
\newblock
\newblock
\urldef\tempurl%
\url{https://www.nber.org/system/files/working_papers/w14808/w14808.pdf}
\showURL{%
\tempurl}


\bibitem[gb(2009)]%
        {gb}
 \bibinfo{year}{2009}\natexlab{}.
\newblock \bibinfo{title}{Good Budget}.
\newblock
\newblock
\urldef\tempurl%
\url{https://goodbudget.com/}
\showURL{%
\tempurl}


\bibitem[pay(2010)]%
        {paytm}
 \bibinfo{year}{2010}\natexlab{}.
\newblock \bibinfo{title}{Paytm}.
\newblock
\newblock
\urldef\tempurl%
\url{https://paytm.com/}
\showURL{%
\tempurl}


\bibitem[roo(2011)]%
        {rooij}
 \bibinfo{year}{2011}\natexlab{}.
\newblock \bibinfo{title}{Financial literacy and stock market participation}.
\newblock
\newblock
\urldef\tempurl%
\url{https://www.sciencedirect.com/science/article/abs/pii/S0304405X11000717}
\showURL{%
\tempurl}


\bibitem[muk(2011)]%
        {mukhopadyay}
 \bibinfo{year}{2011}\natexlab{}.
\newblock \bibinfo{title}{Role of MFIs in Financial Inclusion}.
\newblock
\newblock
\urldef\tempurl%
\url{https://journals.sagepub.com/doi/epdf/10.1177/097492921100300304}
\showURL{%
\tempurl}


\bibitem[spl(2011)]%
        {spl}
 \bibinfo{year}{2011}\natexlab{}.
\newblock \bibinfo{title}{Splitwise}.
\newblock
\newblock
\urldef\tempurl%
\url{https://www.splitwise.com/}
\showURL{%
\tempurl}


\bibitem[goo(2012)]%
        {google_pay}
 \bibinfo{year}{2012}\natexlab{}.
\newblock \bibinfo{title}{GOOGLE PAY}.
\newblock
\newblock
\urldef\tempurl%
\url{https://pay.google.com/about/}
\showURL{%
\tempurl}


\bibitem[llm(2012)]%
        {llm}
 \bibinfo{year}{2012}\natexlab{}.
\newblock \bibinfo{title}{OPTIMAL FINANCIAL KNOWLEDGE AND WEALTH INEQUALITY}.
\newblock
\newblock
\urldef\tempurl%
\url{https://www.nber.org/system/files/working_papers/w18669/w18669.pdf}
\showURL{%
\tempurl}


\bibitem[stu(2012)]%
        {student_debt}
 \bibinfo{year}{2012}\natexlab{}.
\newblock \bibinfo{title}{Student Debt and the Class of 2011}.
\newblock
\newblock
\urldef\tempurl%
\url{https://files.eric.ed.gov/fulltext/ED537338.pdf}
\showURL{%
\tempurl}


\bibitem[ken(2013)]%
        {kenya}
 \bibinfo{year}{2013}\natexlab{}.
\newblock \bibinfo{title}{31-of-kenyas-gdp-is-spent-through-mobile-phones}.
\newblock
\newblock
\urldef\tempurl%
\url{https://qz.com/57504/31-of-kenyas-gdp-is-spent-through-mobile-phones}
\showURL{%
\tempurl}


\bibitem[hon(2013)]%
        {honeybook}
 \bibinfo{year}{2013}\natexlab{}.
\newblock \bibinfo{title}{Honeybook}.
\newblock
\newblock
\urldef\tempurl%
\url{https://www.honeybook.com/}
\showURL{%
\tempurl}


\bibitem[spe(2013)]%
        {spen}
 \bibinfo{year}{2013}\natexlab{}.
\newblock \bibinfo{title}{Spendee}.
\newblock
\newblock
\urldef\tempurl%
\url{https://www.spendee.com/}
\showURL{%
\tempurl}


\bibitem[asi(2014)]%
        {asic}
 \bibinfo{year}{2014}\natexlab{}.
\newblock \bibinfo{title}{Shaping a National Financial Literacy Strategy for 2014–16}.
\newblock
\newblock
\urldef\tempurl%
\url{https://download.asic.gov.au/media/1335644/cp206-published-30-April-2013.pdf}
\showURL{%
\tempurl}


\bibitem[wal(2014)]%
        {wall}
 \bibinfo{year}{2014}\natexlab{}.
\newblock \bibinfo{title}{Wallet - Daily Budget and Profit}.
\newblock
\newblock
\urldef\tempurl%
\url{https://budgetbakers.com/}
\showURL{%
\tempurl}


\bibitem[pho(2015)]%
        {phone_pe}
 \bibinfo{year}{2015}\natexlab{}.
\newblock \bibinfo{title}{Phone Pe}.
\newblock
\newblock
\urldef\tempurl%
\url{https://www.phonepe.com/}
\showURL{%
\tempurl}


\bibitem[bhi(2016)]%
        {bhim}
 \bibinfo{year}{2016}\natexlab{}.
\newblock \bibinfo{title}{BHIM UPI}.
\newblock
\newblock
\urldef\tempurl%
\url{https://www.bhimupi.org.in/}
\showURL{%
\tempurl}


\bibitem[goe(2016)]%
        {goel}
 \bibinfo{year}{2016}\natexlab{}.
\newblock \bibinfo{title}{CYBER-CRIME: A GROWING THREAT TO INDIAN BANKING SECTOR}.
\newblock
\newblock
\urldef\tempurl%
\url{http://data.conferenceworld.in/IFUNA18DEC16/P13-20.pdf}
\showURL{%
\tempurl}


\bibitem[fin(2016)]%
        {fin}
 \bibinfo{year}{2016}\natexlab{}.
\newblock \bibinfo{title}{FinArt App}.
\newblock
\newblock
\urldef\tempurl%
\url{https://www.finart.app/}
\showURL{%
\tempurl}


\bibitem[my(2016)]%
        {my}
 \bibinfo{year}{2016}\natexlab{}.
\newblock \bibinfo{title}{Monefy}.
\newblock
\newblock
\urldef\tempurl%
\url{https://monefy.me/}
\showURL{%
\tempurl}


\bibitem[sno(2016)]%
        {snotes}
 \bibinfo{year}{2016}\natexlab{}.
\newblock \bibinfo{title}{Samsung notes}.
\newblock
\newblock
\urldef\tempurl%
\url{https://www.samsung.com/us/support/owners/app/samsung-notes}
\showURL{%
\tempurl}


\bibitem[ter(2017)]%
        {terrorism}
 \bibinfo{year}{2017}\natexlab{}.
\newblock \showarticletitle{Corruption, black money \& terrorism are festering sores}.
\newblock  (\bibinfo{date}{Aug.} \bibinfo{year}{2017}).
\newblock
\urldef\tempurl%
\url{https://www.thehindu.com/news/resources/demonetisation-of-rs-500-1000-notes-text-of-modis-address-to-the-nation/article16440798.ece}
\showURL{%
\tempurl}


\bibitem[yon(2017)]%
        {yono}
 \bibinfo{year}{2017}\natexlab{}.
\newblock \bibinfo{title}{Yono}.
\newblock
\newblock
\urldef\tempurl%
\url{https://yonobusiness.sbi/}
\showURL{%
\tempurl}


\bibitem[pay(2018)]%
        {payhawk}
 \bibinfo{year}{2018}\natexlab{}.
\newblock \bibinfo{title}{Payhawk}.
\newblock
\newblock
\urldef\tempurl%
\url{https://payhawk.com/}
\showURL{%
\tempurl}


\bibitem[tan(2019)]%
        {tanmay}
 \bibinfo{year}{2019}\natexlab{}.
\newblock \bibinfo{title}{Digital Wallets ‘Turning a Corner’ for Financial Inclusion: A Study of Everyday}.
\newblock
\newblock
\urldef\tempurl%
\url{https://inria.hal.science/hal-02281325/document}
\showURL{%
\tempurl}


\bibitem[2ND(2019)]%
        {2NDLARGEST}
 \bibinfo{year}{2019}\natexlab{}.
\newblock \bibinfo{title}{Due to Jio, India is home to world's 2nd largest internet user base: Report}.
\newblock
\newblock
\urldef\tempurl%
\url{https://www.livemint.com/industry/telecom/due-to-jio-india-is-home-to-world-s-2nd-largest-internet-user-base-report-1560344314812.html}
\showURL{%
\tempurl}


\bibitem[Jio(2019)]%
        {Jio}
 \bibinfo{year}{2019}\natexlab{}.
\newblock \showarticletitle{How the 'Jio effect' brought millions of Indians online and is reshaping Silicon Valley and the internet}.
\newblock  (\bibinfo{date}{Aug.} \bibinfo{year}{2019}).
\newblock
\urldef\tempurl%
\url{https://www.businessinsider.in/how-the-jio-effect-brought-millions-of-indians-online-and-is-reshaping-silicon-valley-and-the-internet/articleshow/70723349.cms}
\showURL{%
\tempurl}


\bibitem[mm(2020)]%
        {mm}
 \bibinfo{year}{2020}\natexlab{}.
\newblock \bibinfo{title}{Money Manager}.
\newblock
\newblock
\urldef\tempurl%
\url{https://www.realbyteapps.com/}
\showURL{%
\tempurl}


\bibitem[rbi(2020)]%
        {rbi-nsfe}
 \bibinfo{year}{2020}\natexlab{}.
\newblock \bibinfo{title}{National Strategy for Financial Education}.
\newblock
\newblock
\urldef\tempurl%
\url{https://rbidocs.rbi.org.in/rdocs/PublicationReport/Pdfs/NSFE016072012.pdf}
\showURL{%
\tempurl}


\bibitem[glo(2021)]%
        {globalfindex}
 \bibinfo{year}{2021}\natexlab{}.
\newblock \bibinfo{title}{The Global Findex Database 2021}.
\newblock
\newblock
\urldef\tempurl%
\url{https://thedocs.worldbank.org/en/doc/4c4fe6db0fd7a7521a70a39ac518d74b-0050062022/original/Findex2021-India-Country-Brief.pdf}
\showURL{%
\tempurl}


\bibitem[sca(2021)]%
        {scam}
 \bibinfo{year}{2021}\natexlab{}.
\newblock \bibinfo{title}{UPI: A TECHNOLOGICAL CHANGE OR A NEW WAY OF SCAM}.
\newblock
\newblock
\urldef\tempurl%
\url{https://3fdef50c-add3-4615-a675-a91741bcb5c0.usrfiles.com/ugd/3fdef5_fdd19946f8934f71a1c184bf603c7a4d.pdf}
\showURL{%
\tempurl}


\bibitem[wor(2022)]%
        {worldbank}
 \bibinfo{year}{2022}\natexlab{}.
\newblock \bibinfo{title}{COVID-19 Drives Global Surge in use of Digital Payments}.
\newblock
\newblock
\urldef\tempurl%
\url{https://www.worldbank.org/en/news/press-release/2022/06/29/covid-19-drives-global-surge-in-use-of-digital-payments}
\showURL{%
\tempurl}


\bibitem[MIN(2022)]%
        {MINT}
 \bibinfo{year}{2022}\natexlab{}.
\newblock \bibinfo{title}{India has over 1.2 bn mobile phone users: I\&B ministry}.
\newblock
\newblock
\urldef\tempurl%
\url{https://www.livemint.com/technology/gadgets/india-has-over-1-2-bn-mobile-phone-users-i-b-ministry-11668610623295.html}
\showURL{%
\tempurl}


\bibitem[mym(2022)]%
        {mymoney}
 \bibinfo{year}{2022}\natexlab{}.
\newblock \bibinfo{title}{Mymoney}.
\newblock
\newblock
\urldef\tempurl%
\url{https://www.mymoneyapp.co/}
\showURL{%
\tempurl}


\bibitem[sca(2023)]%
        {scam_newspaper}
 \bibinfo{year}{2023}\natexlab{}.
\newblock \bibinfo{title}{Almost half of cyber crime cases since 2020 have been linked to UPI}.
\newblock
\newblock
\urldef\tempurl%
\url{https://www.thehindubusinessline.com/data-stories/data-focus/almost-half-of-cyber-crime-cases-since-2020-have-been-linked-to-upi/article67348520.ece}
\showURL{%
\tempurl}


\bibitem[exp(2023)]%
        {experience}
 \bibinfo{year}{2023}\natexlab{}.
\newblock \bibinfo{title}{Designing Mobile Payment Experiences}.
\newblock
\newblock
\urldef\tempurl%
\url{https://books.google.co.in/books?id=29tFBAAAQBAJ&lpg=PR3&dq=upi%20app%20redesign%20&lr&pg=PR3#v=onepage&q&f=false}
\showURL{%
\tempurl}


\bibitem[89.(2023)]%
        {89.5_million}
 \bibinfo{year}{2023}\natexlab{}.
\newblock \showarticletitle{India tops world ranking in digital payments, records 89.5 million transactions in 2022: MyGovIndia}.
\newblock  (\bibinfo{date}{June} \bibinfo{year}{2023}).
\newblock
\urldef\tempurl%
\url{https://www.businessinsider.in/india/news/india-tops-world-ranking-in-digital-payments-records-89-5-million-transactions-in-2022-mygovindia/articleshow/100891853.cms}
\showURL{%
\tempurl}


\bibitem[pho(2023)]%
        {phonepetimesofindia}
 \bibinfo{year}{2023}\natexlab{}.
\newblock \bibinfo{title}{Most popular UPI apps in India in 2023}.
\newblock
\newblock
\urldef\tempurl%
\url{https://timesofindia.indiatimes.com/gadgets-news/most-popular-upi-apps-in-india-in-2023/photostory/105010892.cms}
\showURL{%
\tempurl}


\bibitem[cas(2023)]%
        {cashsave}
 \bibinfo{year}{2023}\natexlab{}.
\newblock \bibinfo{title}{Personal Finance and Budgeting Mobile Application,“CashSave”}.
\newblock
\newblock
\urldef\tempurl%
\url{https://publisher.uthm.edu.my/periodicals/index.php/aitcs/article/view/7621/3930}
\showURL{%
\tempurl}


\bibitem[npc(2023a)]%
        {npci_livemembers}
 \bibinfo{year}{2023}\natexlab{a}.
\newblock \bibinfo{title}{UPI LIVE MEMBERS}.
\newblock
\newblock
\urldef\tempurl%
\url{https://www.npci.org.in/what-we-do/upi/live-members}
\showURL{%
\tempurl}


\bibitem[npc(2023b)]%
        {npci_productstatistics}
 \bibinfo{year}{2023}\natexlab{b}.
\newblock \bibinfo{title}{UPI PRODUCT STATISTICS}.
\newblock
\newblock
\urldef\tempurl%
\url{https://www.npci.org.in/what-we-do/upi/product-statistics}
\showURL{%
\tempurl}


\bibitem[sin(2023)]%
        {singh}
 \bibinfo{year}{2023}\natexlab{}.
\newblock \bibinfo{title}{UPI scams on the rise: Here's how you can protect yourself}.
\newblock
\newblock
\urldef\tempurl%
\url{https://www.businesstoday.in/tech-today/trending/story/upi-scams-on-the-rise-heres-how-you-can-protect-yourself-383512-2023-05-30}
\showURL{%
\tempurl}


\bibitem[eld(2024)]%
        {elderly}
 \bibinfo{year}{2024}\natexlab{}.
\newblock \bibinfo{title}{Analysing accessibility in e-wallet apps for the elderly}.
\newblock
\newblock
\urldef\tempurl%
\url{10.1504/IJBIS.2024.140434}
\showURL{%
\tempurl}


\bibitem[bus(2024)]%
        {businesstoday}
 \bibinfo{year}{2024}\natexlab{}.
\newblock \bibinfo{title}{Digital payment frauds surge in India as UPI transactions skyrocket: RBI report}.
\newblock
\newblock
\urldef\tempurl%
\url{https://www.businesstoday.in/technology/news/story/digital-payment-frauds-surge-in-india-as-upi-transactions-skyrocket-rbi-report-431695-2024-06-01}
\showURL{%
\tempurl}


\bibitem[pho(2024)]%
        {phonepe}
 \bibinfo{year}{2024}\natexlab{}.
\newblock \bibinfo{title}{Exclusive: India to again delay caps on UPI payments market share}.
\newblock
\newblock
\urldef\tempurl%
\url{https://www.reuters.com/business/finance/india-delay-payments-market-cap-helping-walmart-backed-phonepe-google-pay-2024-05-09/}
\showURL{%
\tempurl}


\bibitem[fig(2024)]%
        {figma}
 \bibinfo{year}{2024}\natexlab{}.
\newblock \bibinfo{title}{FIGMA}.
\newblock
\newblock
\urldef\tempurl%
\url{https://www.figma.com/}
\showURL{%
\tempurl}


\bibitem[tru(2024)]%
        {truecaller}
 \bibinfo{year}{2024}\natexlab{}.
\newblock \bibinfo{title}{How does spam detection happen on the Truecaller app?}
\newblock
\newblock
\urldef\tempurl%
\url{https://docs.truecaller.com/truecaller-for-business/spam-management/how-does-spam-detection-happen-on-the-truecaller-app}
\showURL{%
\tempurl}


\bibitem[sca(2024)]%
        {scammers}
 \bibinfo{year}{2024}\natexlab{}.
\newblock \bibinfo{title}{Why both businesses and scammers love India's payment system}.
\newblock
\newblock
\urldef\tempurl%
\url{https://www.bbc.com/news/articles/c288m1km01po}
\showURL{%
\tempurl}


\bibitem[Alkhowaiter(2020)]%
        {gulf_countries}
\bibfield{author}{\bibinfo{person}{Wassan~Abdullah Alkhowaiter}.} \bibinfo{year}{2020}\natexlab{}.
\newblock \showarticletitle{Digital payment and banking adoption research in Gulf countries: A systematic literature review}.
\newblock \bibinfo{journal}{\emph{International Journal of Information Management}}  \bibinfo{volume}{53} (\bibinfo{year}{2020}), \bibinfo{pages}{102102}.
\newblock
\showISSN{0268-4012}
\urldef\tempurl%
\url{https://doi.org/10.1016/j.ijinfomgt.2020.102102}
\showDOI{\tempurl}


\bibitem[Asongu and Boateng(2018)]%
        {chineseqr_2}
\bibfield{author}{\bibinfo{person}{Simplice Asongu} {and} \bibinfo{person}{Agyenim Boateng}.} \bibinfo{year}{2018}\natexlab{}.
\newblock \showarticletitle{Introduction to Special Issue: Mobile Technologies and Inclusive Development in Africa}.
\newblock \bibinfo{journal}{\emph{Journal of African Business}} \bibinfo{volume}{19}, \bibinfo{number}{3} (\bibinfo{year}{2018}), \bibinfo{pages}{297--301}.
\newblock
\urldef\tempurl%
\url{https://doi.org/10.1080/15228916.2018.1481307}
\showDOI{\tempurl}
\showeprint{https://doi.org/10.1080/15228916.2018.1481307}


\bibitem[Asongu and Odhiambo(2019)]%
        {chineseqr_3}
\bibfield{author}{\bibinfo{person}{Simplice Asongu} {and} \bibinfo{person}{Nicholas Odhiambo}.} \bibinfo{year}{2019}\natexlab{}.
\newblock \showarticletitle{Mobile banking usage, quality of growth, inequality and poverty in developing countries}.
\newblock \bibinfo{journal}{\emph{Information Development}}  \bibinfo{volume}{35} (\bibinfo{date}{March} \bibinfo{year}{2019}), \bibinfo{pages}{303--318}.
\newblock
\urldef\tempurl%
\url{https://doi.org/10.1177/0266666917744006}
\showDOI{\tempurl}


\bibitem[Davis(1989)]%
        {tam1989}
\bibfield{author}{\bibinfo{person}{Fred~D. Davis}.} \bibinfo{year}{1989}\natexlab{}.
\newblock \showarticletitle{Perceived Usefulness, Perceived Ease of Use, and User Acceptance of Information Technology}.
\newblock \bibinfo{journal}{\emph{MIS Quarterly}} \bibinfo{volume}{13}, \bibinfo{number}{3} (\bibinfo{year}{1989}), \bibinfo{pages}{319--340}.
\newblock
\showISSN{02767783, 21629730}
\urldef\tempurl%
\url{http://www.jstor.org/stable/249008}
\showURL{%
\tempurl}


\bibitem[Ferreira and Perry(2019)]%
        {chineseqr_24}
\bibfield{author}{\bibinfo{person}{Jennifer Ferreira} {and} \bibinfo{person}{Mark Perry}.} \bibinfo{year}{2019}\natexlab{}.
\newblock \showarticletitle{From Transactions to Interactions: Social Considerations for Digital Money}.
\newblock  (\bibinfo{year}{2019}), \bibinfo{pages}{121--133}.
\newblock
\showISBNx{978-3-030-02330-0}
\urldef\tempurl%
\url{https://doi.org/10.1007/978-3-030-02330-0_8}
\showDOI{\tempurl}


\bibitem[Fran~Bennett(2009)]%
        {couplesharemoney}
\bibfield{author}{\bibinfo{person}{Capitolina Diaz Bjorn~Hallerod Fran~Bennett, Janet~Stocks}.} \bibinfo{year}{2009}\natexlab{}.
\newblock \showarticletitle{Modern Couples, Sharing Money, Sharing Life}.
\newblock \bibinfo{journal}{\emph{Feminist Economics}} \bibinfo{volume}{15}, \bibinfo{number}{2} (\bibinfo{year}{2009}), \bibinfo{pages}{120--125}.
\newblock
\showISBNx{978-0-2305-1702-8}
\urldef\tempurl%
\url{https://doi.org/10.1080/13545700802698605}
\showDOI{\tempurl}


\bibitem[Ghatak(2017)]%
        {Misleadingdichotomy}
\bibfield{author}{\bibinfo{person}{Aditi~Roy Ghatak}.} \bibinfo{year}{2017}\natexlab{}.
\newblock \showarticletitle{Misleading dichotomy}.
\newblock  (\bibinfo{year}{2017}).
\newblock
\urldef\tempurl%
\url{https://www.dandc.eu/en/article/indias-informal-sector-backbone-economy}
\showURL{%
\tempurl}


\bibitem[Hassan and Shukur(2022)]%
        {electronics}
\bibfield{author}{\bibinfo{person}{Md~Arif Hassan} {and} \bibinfo{person}{Zarina Shukur}.} \bibinfo{year}{2022}\natexlab{}.
\newblock \showarticletitle{Device Identity-Based User Authentication on Electronic Payment System for Secure E-Wallet Apps}.
\newblock \bibinfo{journal}{\emph{Electronics}} \bibinfo{volume}{11}, \bibinfo{number}{1} (\bibinfo{year}{2022}).
\newblock
\showISSN{2079-9292}
\urldef\tempurl%
\url{https://doi.org/10.3390/electronics11010004}
\showDOI{\tempurl}


\bibitem[He et~al\mbox{.}(2023)]%
        {chineseqr}
\bibfield{author}{\bibinfo{person}{Changyang He}, \bibinfo{person}{Lu He}, \bibinfo{person}{Zhicong Lu}, {and} \bibinfo{person}{Bo Li}.} \bibinfo{year}{2023}\natexlab{}.
\newblock \showarticletitle{"I Have to Use My Son's QR Code to Run the Business": Unpacking Senior Street Vendors' Challenges in Mobile Money Collection in China}.
\newblock \bibinfo{journal}{\emph{Proc. ACM Hum.-Comput. Interact.}} \bibinfo{volume}{7}, \bibinfo{number}{CSCW1}, Article \bibinfo{articleno}{60} (\bibinfo{date}{apr} \bibinfo{year}{2023}), \bibinfo{numpages}{28}~pages.
\newblock
\urldef\tempurl%
\url{https://doi.org/10.1145/3579493}
\showDOI{\tempurl}


\bibitem[Kameswaran and Hulikal~Muralidhar(2019)]%
        {Cash-Digital-Payments-and-Accessibility-A-Case-Study-from-India}
\bibfield{author}{\bibinfo{person}{Vaishnav Kameswaran} {and} \bibinfo{person}{Srihari Hulikal~Muralidhar}.} \bibinfo{year}{2019}\natexlab{}.
\newblock \showarticletitle{Cash, Digital Payments and Accessibility: A Case Study from Metropolitan India}.
\newblock \bibinfo{journal}{\emph{Proc. ACM Hum.-Comput. Interact.}} \bibinfo{volume}{3}, \bibinfo{number}{CSCW}, Article \bibinfo{articleno}{97} (\bibinfo{date}{nov} \bibinfo{year}{2019}), \bibinfo{numpages}{23}~pages.
\newblock
\urldef\tempurl%
\url{https://doi.org/10.1145/3359199}
\showDOI{\tempurl}


\bibitem[Kaye et~al\mbox{.}(2014)]%
        {moneymatters}
\bibfield{author}{\bibinfo{person}{Joseph Kaye}, \bibinfo{person}{Mary McCuistion}, \bibinfo{person}{Rebecca Gulotta}, {and} \bibinfo{person}{David Shamma}.} \bibinfo{year}{2014}\natexlab{}.
\newblock \showarticletitle{Money talks: Tracking personal finances}.
\newblock \bibinfo{journal}{\emph{Conference on Human Factors in Computing Systems - Proceedings}} (\bibinfo{date}{04} \bibinfo{year}{2014}).
\newblock
\showISBNx{978-1-4503-2473-1}
\urldef\tempurl%
\url{https://doi.org/10.1145/2556288.2556975}
\showDOI{\tempurl}


\bibitem[Lewis and Perry(2019)]%
        {chineseqr_52}
\bibfield{author}{\bibinfo{person}{Makayla Lewis} {and} \bibinfo{person}{Mark Perry}.} \bibinfo{year}{2019}\natexlab{}.
\newblock \showarticletitle{Follow the Money: Managing Personal Finance Digitally}. In \bibinfo{booktitle}{\emph{Proceedings of the 2019 CHI Conference on Human Factors in Computing Systems}} (Glasgow, Scotland, UK) \emph{(\bibinfo{series}{CHI '19})}. \bibinfo{publisher}{Association for Computing Machinery}, \bibinfo{address}{New York, NY, USA}, \bibinfo{pages}{1--14}.
\newblock
\showISBNx{9781450359702}
\urldef\tempurl%
\url{https://doi.org/10.1145/3290605.3300620}
\showDOI{\tempurl}


\bibitem[Lusardi and Mitchell(2014)]%
        {mitchell_lusardi}
\bibfield{author}{\bibinfo{person}{Annamaria Lusardi} {and} \bibinfo{person}{Olivia~S. Mitchell}.} \bibinfo{year}{2014}\natexlab{}.
\newblock \showarticletitle{The Economic Importance of Financial Literacy: Theory and Evidence}.
\newblock \bibinfo{journal}{\emph{Journal of Economic Literature}} \bibinfo{volume}{52}, \bibinfo{number}{1} (\bibinfo{date}{March} \bibinfo{year}{2014}), \bibinfo{pages}{5--44}.
\newblock
\urldef\tempurl%
\url{https://doi.org/10.1257/jel.52.1.5}
\showDOI{\tempurl}


\bibitem[Mistry(2023)]%
        {riseofupi}
\bibfield{author}{\bibinfo{person}{Mehul Mistry}.} \bibinfo{year}{2023}\natexlab{}.
\newblock \showarticletitle{The Rise of UPI: Transforming the way Indians transact}.
\newblock  (\bibinfo{date}{July} \bibinfo{year}{2023}).
\newblock
\urldef\tempurl%
\url{https://timesofindia.indiatimes.com/blogs/voices/the-rise-of-upi-transforming-the-way-indians-transact/}
\showURL{%
\tempurl}


\bibitem[Pal et~al\mbox{.}(2018a)]%
        {digitalDiscontent}
\bibfield{author}{\bibinfo{person}{Joyojeet Pal}, \bibinfo{person}{Priyank Chandra}, \bibinfo{person}{Vaishnav Kameswaran}, \bibinfo{person}{Aakanksha Parameshwar}, \bibinfo{person}{Sneha Joshi}, {and} \bibinfo{person}{Aditya Johri}.} \bibinfo{year}{2018}\natexlab{a}.
\newblock \showarticletitle{Digital Payment and Its Discontents: Street Shops and the Indian Government's Push for Cashless Transactions}. In \bibinfo{booktitle}{\emph{Proceedings of the 2018 CHI Conference on Human Factors in Computing Systems}} (<conf-loc>, <city>Montreal QC</city>, <country>Canada</country>, </conf-loc>) \emph{(\bibinfo{series}{CHI '18})}. \bibinfo{publisher}{Association for Computing Machinery}, \bibinfo{address}{New York, NY, USA}, \bibinfo{pages}{1–13}.
\newblock
\showISBNx{9781450356206}
\urldef\tempurl%
\url{https://doi.org/10.1145/3173574.3173803}
\showDOI{\tempurl}


\bibitem[Pal et~al\mbox{.}(2018b)]%
        {chineseqr_68}
\bibfield{author}{\bibinfo{person}{Joyojeet Pal}, \bibinfo{person}{Priyank Chandra}, \bibinfo{person}{Vaishnav Kameswaran}, \bibinfo{person}{Aakanksha Parameshwar}, \bibinfo{person}{Sneha Joshi}, {and} \bibinfo{person}{Aditya Johri}.} \bibinfo{year}{2018}\natexlab{b}.
\newblock \showarticletitle{Digital Payment and Its Discontents: Street Shops and the Indian Government's Push for Cashless Transactions}. \bibinfo{pages}{1--13}.
\newblock
\urldef\tempurl%
\url{https://doi.org/10.1145/3173574.3173803}
\showDOI{\tempurl}


\bibitem[Shiva(2017)]%
        {sudheer}
\bibfield{author}{\bibinfo{person}{Hemanth~Kumar Shiva}.} \bibinfo{year}{2017}\natexlab{}.
\newblock \showarticletitle{EPRA International Journal of Economic and Business Review A STUDY ON SECURED CASHLESS ECONOMY IN INDIA WITH REFERENCE TO INITIATIVES TAKEN BY RESERVE BANK OF INDIA (RBI) TOWARDS DIGITAL PAYMENT}.
\newblock \bibinfo{journal}{\emph{EPRA International Journal of Economic and Business Review}}  \bibinfo{volume}{5} (\bibinfo{date}{09} \bibinfo{year}{2017}), \bibinfo{pages}{21--27}.
\newblock
\urldef\tempurl%
\url{https://doi.org/10.5281/zenodo.4679310}
\showDOI{\tempurl}


\bibitem[Sreenu(2020)]%
        {cashless_economy_growth}
\bibfield{author}{\bibinfo{person}{Nenavath Sreenu}.} \bibinfo{year}{2020}\natexlab{}.
\newblock \showarticletitle{Cashless Payment Policy and Its Effects on Economic Growth of India: An Exploratory Study}.
\newblock \bibinfo{journal}{\emph{ACM Trans. Manage. Inf. Syst.}} \bibinfo{volume}{11}, \bibinfo{number}{3}, Article \bibinfo{articleno}{15} (\bibinfo{date}{aug} \bibinfo{year}{2020}), \bibinfo{numpages}{10}~pages.
\newblock
\showISSN{2158-656X}
\urldef\tempurl%
\url{https://doi.org/10.1145/3391402}
\showDOI{\tempurl}


\bibitem[Srivastava(2022)]%
        {financial_inclusion}
\bibfield{author}{\bibinfo{person}{Anushree Srivastava}.} \bibinfo{year}{2022}\natexlab{}.
\newblock \showarticletitle{Digital Financial Inclusion: A Study to Find the Role of Financial and Digital Literacy in Achieving It}.
\newblock \bibinfo{journal}{\emph{Int. J. Innov. Digit. Econ.}} \bibinfo{volume}{13}, \bibinfo{number}{1} (\bibinfo{date}{jul} \bibinfo{year}{2022}), \bibinfo{pages}{1–12}.
\newblock
\showISSN{1947-8305}
\urldef\tempurl%
\url{https://doi.org/financial_inclusion}
\showDOI{\tempurl}


\bibitem[Surendran et~al\mbox{.}(2018)]%
        {ieee}
\bibfield{author}{\bibinfo{person}{Sudheer Surendran}, \bibinfo{person}{B. Sivaselvan}, {and} \bibinfo{person}{C Oswald}.} \bibinfo{year}{2018}\natexlab{}.
\newblock \showarticletitle{Emergent User Design Framework for E Payment Mobile Application}. In \bibinfo{booktitle}{\emph{2018 International Conference on Computer, Communication, and Signal Processing (ICCCSP)}}. \bibinfo{pages}{1--6}.
\newblock
\urldef\tempurl%
\url{https://doi.org/10.1109/ICCCSP.2018.8452855}
\showDOI{\tempurl}


\bibitem[Vashistha et~al\mbox{.}(2019)]%
        {chineseqr_97}
\bibfield{author}{\bibinfo{person}{Aditya Vashistha}, \bibinfo{person}{Richard Anderson}, {and} \bibinfo{person}{Shrirang Mare}.} \bibinfo{year}{2019}\natexlab{}.
\newblock \showarticletitle{Examining the Use and Non-Use of Mobile Payment Systems for Merchant Payments in India}. In \bibinfo{booktitle}{\emph{Proceedings of the 2nd ACM SIGCAS Conference on Computing and Sustainable Societies}} (Accra, Ghana) \emph{(\bibinfo{series}{COMPASS '19})}. \bibinfo{publisher}{Association for Computing Machinery}, \bibinfo{address}{New York, NY, USA}, \bibinfo{pages}{1--12}.
\newblock
\showISBNx{9781450367141}
\urldef\tempurl%
\url{https://doi.org/10.1145/3314344.3332499}
\showDOI{\tempurl}


\bibitem[Vines et~al\mbox{.}(2011)]%
        {oldies}
\bibfield{author}{\bibinfo{person}{John Vines}, \bibinfo{person}{Mark Blythe}, \bibinfo{person}{Paul Dunphy}, {and} \bibinfo{person}{Andrew Monk}.} \bibinfo{year}{2011}\natexlab{}.
\newblock \showarticletitle{Eighty Something: Banking for the Older Old}. In \bibinfo{booktitle}{\emph{Proceedings of the 25th BCS Conference on Human-Computer Interaction}} (Newcastle-upon-Tyne, United Kingdom) \emph{(\bibinfo{series}{BCS-HCI '11})}. \bibinfo{publisher}{BCS Learning \& Development Ltd.}, \bibinfo{address}{Swindon, GBR}, \bibinfo{pages}{64–73}.
\newblock


\end{thebibliography}


\appendix \label{sec:appendix}

\section{\textbf{PHASE 1 : SURVEY QUESTIONS}}\label{appendix:phase_1_survey}

1. What is your age? (Open-ended)

2. What is your profession? (Multiple Choice) 
   (Options: Student, Working Professional, Business, Others)

3. What is your gender? (Multiple Choice) 
   (Options: Male, Female, Non-binary, Prefer not to say)

4. Have you used UPI as the payment method? (Yes/No)

5. If yes, which year did you start using it? (Open-ended)

6. How often do you use UPI for making payments? (Multiple Choice) 
   (Options: Daily, Weekly, Monthly, Occasionally, Rarely)

7. Have you noticed any changes in your spending habits since you started using UPI? (Multiple Choice) 
   (Options: Increased Spending, Decreased Spending, No Change)

8. On a scale of 1 to 5, rate the convenience you find with UPI for making payments. (Likert Scale)

9. How satisfied are you with your overall experience using UPI for payments? (Likert Scale) 
   (1 = Not Satisfied at all, 5 = Extremely Satisfied)

10. On average, how much do you spend daily using UPI? (Open-ended)

11. Have you found any change in your spending ever since using UPI? (Yes/No)

12. Do you feel that using UPI affects your ability to stick to your budget? (Likert Scale) 
    (Options: Strongly Disagree, Disagree, Neutral, Agree, Strongly Agree)

13. Have you experienced any instances where you have exceeded your budget or spent more than intended due to UPI usage? (Yes/No)

14. How often do you check your transaction history or keep track of your UPI spending? (Multiple Choice) 
    (Options: Rarely, Occasionally, Frequently, Very Frequently, Regularly)

15. How do you manage/track your expenses? (Open-ended)

16. Additional Comments: Do you want to share anything about your experience with UPI and its impact on your spending behavior? (Open-ended)

17. Will you be available for online interviews if called? (Yes/No)

18. If yes, provide your email address: (Open-ended)

19. If yes, provide your phone number: (Open-ended)

\section{\textbf{PHASE 1 : INTERVIEW QUESTIONS}}\label{appendix:phase_1_interviews}

\textbf{Demographic Information}

1. What is your age? (Open-ended)

2. What is your gender? (Multiple Choice) 
   (Options: Male, Female, Non-binary, Prefer not to say)

3. What is your occupation? (Open-ended)

\textbf{Usage of UPI}

4. Have you used UPI as a payment method? (Yes/No)

5. If yes, when did you start using it? (Open-ended)

6. How often do you use UPI to make payments? (Multiple Choice) 
   (Options: Daily, Weekly, Monthly, Occasionally, Rarely)

7. Can you briefly describe your experience with using UPI to make transactions? (Open-ended)

8. Have you noticed any changes in your spending habits since you started using UPI? (Multiple Choice) 
   (Options: Increased Spending, Decreased Spending, No Change)

9. If yes, could you elaborate on the changes you have observed in your spending habits after adopting UPI? (Open-ended)

\textbf{Perception of UPI Convenience}

10. On a scale of 1 to 5, rate the convenience you find with UPI for making payments. (Likert Scale) 
    (1 = Not Convenient at all, 5 = Extremely Convenient)

11. In what ways do you perceive UPI to be convenient/inconvenient for you? (Open-ended)

12. On a scale of 1 to 5, how would you rate the convenience of using UPI compared to traditional payment methods like cash? (Likert Scale) 
    (1 = Not Convenient at all, 5 = Extremely Convenient)

13. How satisfied are you with your overall experience of using UPI for payments? (Likert Scale) 
    (1 = Not Satisfied at all, 5 = Extremely Satisfied)

\textbf{Income Information}

14. What is your approximate annual income range? (Open-ended)

\textbf{Spending Behavior}

15. On average, what is the transaction amount when you use UPI? (Open-ended)

16. On average, how much do you spend daily using UPI? (Open-ended)

17. On average, how much do you spend monthly using UPI? (Open-ended)

18. Where do you spend your money? (Open-ended)

19. Does the spending amount include only UPI transactions or all spending in general? (Open-ended)

20. Have you found any change in your spending ever since using UPI? (Yes/No)

21. If yes, could you specify the reasons for the increased spending after adopting UPI? (Open-ended)

22. Do you feel that using UPI affects your ability to stick to your budget? (Likert Scale) 
    (Options: Strongly Disagree, Disagree, Neutral, Agree, Strongly Agree)

23. How often do you check your transaction history or keep track of your UPI spending? (Multiple Choice) 
    (Options: Rarely, Occasionally, Frequently, Very Frequently, Regularly)

24. Have you experienced any instances where you have exceeded your budget or spent more than intended due to UPI usage? (Yes/No)

25. In your opinion, does using UPI encourage you to spend more money overall? (Likert Scale) 
    (Options: Strongly Disagree, Disagree, Neutral, Agree, Strongly Agree)

\textbf{Factors Influencing Spending Behavior}

26. Besides UPI usage, do other factors have influenced your spending behavior? (Yes/No)

27. If yes, please specify those factors. (Open-ended)

28. Are you aware of any psychological factors influencing your spending behavior when using UPI? If yes, please elaborate. (Open-ended)

\textbf{Attitudes Towards Financial Management}

29. How would you describe your level of financial discipline? (Likert Scale) 
    (1 to 5)

30. Do you feel that using UPI has made managing your finances easier or more difficult? (Multiple Choice) 
    (Options: Easier, More Difficult, No Change)

31. How do you rate the statement "Do you feel that UPI provides you with a clear understanding of your spending patterns"? (Likert Scale) 
    (Options: Strongly Disagree, Disagree, Neutral, Agree, Strongly Agree)

32. How do you rate the statement "UPI has simplified your financial management"? (Likert Scale) 
    (Options: Strongly Disagree, Disagree, Neutral, Agree, Strongly Agree)

\textbf{Additional Comments} 

33. Is there anything else you would like to share about your experience with UPI and its impact on your spending behavior? (Open-ended)

\section{\textbf{PHASE 2 : QUESTIONNAIRE }}\label{appendix:phase_2_questionnaire}

1. Age:
(Open-ended)

2. Profession (Options: Student, Working Professional, Business, Others):
(MCQ)

3. Gender (Options: Male, Female, Others):
(MCQ)

4. Have you noticed any changes in your spending habits since you started using UPI? (Options: Increased Spending, Decreased Spending, Prefer not to disclose):
(MCQ)

5. Which payment app do you use most often? (Options: PhonePe, Google Pay, Paytm, BHIM UPI, Others):
(MCQ)

\textbf{Expense Tracker}

6. Do you believe adding an expense tracker to financial apps like PhonePe, Paytm, etc., would be beneficial? (Options: Not useful at all, Not useful, Neutral, Useful, Extremely Useful):
(Likert scale)

7. Do you find it helpful to view expenses categorized by month and type (e.g., food, travel) with a day-wise breakdown and budget assigned to each category? (Options: Yes, it is very helpful; Yes, it is helpful but I don't think it should be implemented; No, it is not helpful; No, it is confusing and unnecessary; Not able to answer):
(MCQ)

8. Please provide any suggestions you have for improving the expense tracker:
(Open-ended)

\textbf{Balance Display }

9. Do you think it is good to see your balance before and after making a payment? (Options: Yes, it is very helpful to understand and spend less; No, it does not add value and could be distracting; Not able to answer):
(MCQ)

10. Does seeing your balance before payment help you reconsider large payments? (Options: Yes; No; Sometimes; Not able to answer):
(MCQ)

\textbf{Financial Management  }

11. Do you find the one-liner quotes about financial management, budgeting techniques, and related financial facts valuable? (Options: Yes, they can help me spend less and manage my finances better; Yes, they are informative but may not significantly impact my spending habits; No, they could be distracting and unlikely to influence my spending behavior; Not able to answer):
(MCQ)

\textbf{Budget Limitations  }

12. Do you think setting budget limits for specific categories (like food, travel) will help you manage your spending? (Options: Yes, it will help me spend less and manage my finances better; Yes, it might help, but I’m not sure how effective it will be; No, it likely won’t make a significant difference in my spending habits; No, I don’t think it will help with managing spending or reducing guilt; Not able to answer):
(MCQ)

13. How often do you feel guilt or regret about overspending? (Options: Never, Rarely, Sometimes, Often, Always):
(MCQ)

14. Describe a scenario where setting a budget limit for a specific category (e.g., food, travel) would be particularly beneficial for you:
(Open-ended)

\textbf{Payment Type Selection} 

15. Do you find it useful to select and label the type of payment (e.g., food, travel) either before or after proceeding with a transaction? (Options: Extremely useful, it helps me categorize and manage expenses better; Very useful, but it could be streamlined; Moderately useful, though it could be improved; Slightly useful, but it feels like unnecessary jargon; Not useful at all, it seems like a time waste; Not able to answer):
(MCQ)

16. How can the feature to assign and label payment types be made more intuitive or useful? (Options: Simplify the labeling process to reduce unnecessary steps, Automatically categorize payments based on transaction history, Allow users to quickly select from frequently used categories, Provide an option to skip labeling for quick transactions, Use icons or visuals to represent different payment categories, Implement a drag-and-drop feature for categorizing expenses, Enable voice commands to assign and label payments, Provide personalized suggestions based on previous spending behavior, Not able to answer):
(Multiple select)

\textbf{Saving Streak Feature} 

17. Would you be motivated to save more if there were a saving streak feature that rewards you based on how much you save in different categories and overall? (Options: Yes, the rewards would motivate me to save more; No, I don’t think rewards would make a difference in my saving habits; Maybe, it depends on the type and value of the reward; Not able to answer):
(MCQ)

18. What additional gamification features would you like to see on the app? (Options: Boosted cashback rewards for specific types of transactions or merchants, Incentives for making consecutive payments or transactions, Competitions that reward users for timely bill payments, Challenges that encourage sticking to budgeting goals with rewards, Exclusive, personalized offers based on spending patterns, Interactive scenarios to improve financial knowledge with immediate rewards, Random bonuses or discounts awarded for regular transactions, Not able to answer):
(Multiple select)

\textbf{Delay Feature}

19. Would you find it helpful to have a timer (e.g., 8-10 seconds) before completing a large payment, allowing you time to reconsider and confirm if you really want to proceed? (Options: Yes, it would help me make more deliberate decisions and avoid impulsive spending; Yes, it might be useful, but I’m not sure if I’d use it frequently; No, it seems unnecessary and could be an inconvenience; No, I don’t think it would have an impact on my spending decisions; Not able to answer):
(MCQ)

20. What do you consider a large payment in terms of rupees?
(Open-ended)

\textbf{Follow up Questions}

21. Will you be available for online interviews?
(Yes/No)

22. If yes, provide your email address:
(Open-ended)

23. If yes, provide your phone number:
(Open-ended)



\end{document}